\pdfoutput=1

\documentclass[aps,showpacs,preprintnumbers,amsmath,amssymb,
eqsecnum, twocolumn, tightenlines
]{revtex4}

\usepackage{graphicx}

\newcommand{\be}{\begin{eqnarray}}
\newcommand{\ee}{\end{eqnarray}}

 \newcommand{\gsim}{\mathrel{\hbox{\rlap{\lower.55ex \hbox {$\sim$}}
                   \kern-.3em \raise.4ex \hbox{$>$}}}}
\newcommand{\lsim}{\mathrel{\hbox{\rlap{\lower.55ex \hbox {$\sim$}}
                   \kern-.3em \raise.4ex \hbox{$<$}}}}

\newcommand{\ba}{\begin{eqnarray}}
\newcommand{\ea}{\end{eqnarray}}

\def\roughly#1{\mathrel{\raise.3ex\hbox{$#1$\kern-.75em%
\lower1ex\hbox{$\sim$}}}}
\def\lsim{\roughly<}
\def\gsim{\roughly>}

\begin{document}


\title{Sound Waves from Quenched Jets}
\author {Vladimir Khachatryan and Edward Shuryak}
\address {Department of Physics and Astronomy, State University of New York,
Stony Brook, NY 11794}
\date{\today}

\begin{abstract}
Heavy ion collisions at RHIC/LHC energies are well described by the (nearly ideal) 
hydrodynamics. Last year this success has been extended to higher angular harmonics,
$v_n,n=3..9$ induced by  initial-state perturbations, in analogy to cosmic microwave 
background fluctuations. 
Here we  use hydrodynamics to study 
sound propagation emitted by quenched jets.
 We use the so called ``geometric acoustics" 
to follow the sound propagation , on top of the expanding fireball.  The conical waves,  known as 
''Mach cones'', turn out to be strongly
distorted. We show that large radial flow makes  the observed particle
spectra to be determined mostlly by the vicinity of  their intersection with the fireball's space-like and time-like
 freezeout surfaces. We further show how the waves modify the freezeout surfaces and spectra. We end up comparing
 our calculations to the two-particle correlation functions at  RHIC, while emphasizing that studies of dijet events observed at LHC  
should provide much better test of our theory.
\end{abstract}
\maketitle

\section{Introduction}
\subsection{Hydrodynamics and sounds}
Jet-induced correlations have short and remarkable history. Two-hadron correlation functions
measured in AuAu collisions at RHIC have revealed structures
known as ``ridge" and ``shoulders" (about 2 rad away from trigger) These structures were originally
belived to be associated with jets, but have recently been explained as hydrodynamical
``harmonic flows", basically a sound circles created by initial local perturbations. This subject has been discussed extensively
in three papers   \cite{I,II,III} coauthored by one of us, which have 
 very extensive introduction and references, to which the reader may refer.
The first paper of this series \cite{I} has been
discussing  some generic features of the sound propagation and possible 
role of electric (dual-magnetohydrodynamical) corona. The second paper \cite{II} has included more detailed 
studies of the $phases$ of the perturbations and possible ways of their  experimental/theoretical studies.
The 
third paper \cite{III} used  the so called {\em Gubser flow}, which allows analytic treatment of all angular harmonics,
with and without viscosity. (One element of this last paper,
the modification of the freezeout surface by the
 sound waves, will be extensively used at the end of this work.)
 While thermo/hydrodynamics was traditionally expected to work 
for majority of soft particles only, it describes well these harmonic flows including 
the so called ``intermediate transverse momenta" region, $p_t=2-3\, GeV$. Such particles are 
rare, about one such particle per event among thousands. So, once again,  one finds that thermo/hydrodynamics is very robust and can describe not only the  behavior of an average particle, but also the shape of
rather far-reaching tails of thermal/hydro spectra. 

These developments had  (perhaps temporarily) take attention away from the issue of jets
and their interaction with the medium. However, as soon as the trigger hadron has $p_t$ above the intermediate $p_t$ domain,
say $p_t> 6\, GeV$, 
it gets out of reach for collective flows and  comes predominantly from hard collisions, or ``jets". So, the experimental input to  
 the present paper are  two-particle correlations in which the trigger is ``hard", while one (or more) 
associate particles are in the hydro domain,  $p_t=1-3\, GeV$ (hard-intermediate correlations, for short).

 The idea, that once the
energy is deposited into the medium by a jet will be resulting in
sound perturbations in the shape of  the Mach cone, has been
proposed in Refs \cite{Casalderrey-Solana:2004qm,Horst}.
(It may resemble similar idea discussed in 1970's for non-relativistic
nuclear collisions. However it did not work, because nuclear matter
is dilute and not a particularly good liquid, unlike sQGP under discussion now.) 
In the context of strongly coupled QGP, the problem has been
addressed in the framework of AdS/CFT. As detailed by Chesler and
Yaffe \cite{Chesler:2007an}, and Gubser et al
\cite{Gubser:2008yx}, the stress tensor solution obtained by
holographic imaging is in remarkably good agreement with
hydrodynamical solution detailed in
\cite{Casalderrey-Solana:2004qm}, displaying the Mach cone in all
its glory.
 Since the
time-averaged sound velocity over the QGP, mixed and hadronic
phases is
$
<c_{s}>\,\approx\,0.4\,,
$
the expected Mach cone angle
\be %
\theta_{M} = arccos \left({<c_{s}> \over v_{jet} } \right) \approx
1.159\,rad \approx 66.4^{0} \label{eqn_Mach}
\ee %
roughly matches the angular positions of the ``shoulders" at 
\,$\phi\approx \pi \pm \theta_{M}$ in the correlation functions. 
 However, 
as the reader will see from what follows, 
 a flow normal to the Mach cone
is only one of several  geometrical issues involved.
Another, equally important effect, is the radial flow of extra matter produced by the wave.
As we will see, it is defined mostly at the point where the sound cones intersect with the time-like and space-like freezeout
surfaces. 

One more geometrical effects come from the interplay between the jet ``stopping distance" and the actual size of the fireball
(along the jet direction). Obviously, when the two are equal, the amplitude of the wave is at its maximum.
This effect  has been discussed before 
 in direct hydrodynamical
simulations of the cones, such as, e.g., that done by Betz et al
\cite{Betz:2010qh}, which we, to some extend, will follow.

   The main technical difficulty of the problem is in correct treatment of
the sound propagation  on top of the expanding fireball.  This problem has been first
discussed by Casalderrey-Solana and Shuryak in
\cite{CasalderreySolana:2005rf} which modeled the fireball
expansion  by a Hubble-like overall expansion of the metric, the
same  Friedman-Robertson-Walker metric as used for Big Bang.
(Of course, it creates new space rather than expanding, so it is not literally corresponding to the problem at hand, it was just a technical first step.)
The physics focus of that paper was the effect of time-dependent sound
velocity, especially if the phase transition is of the 1st order and 
can vanish at some interval of temperature $T$. The interesting finding 
was a creation of secondary -- and convergent -- sound waves. This
idea was further discussed in \cite{I} in connection with the
``soft ridge" issue. However,  if the current lattice
data on the speed of sound are correct, the calculated effect of the
reflected wave was shown to be  too small to explain the ``hard ridge".

In this paper we developed a different method based on ``geometric acoustics.

Let us now explain the structure of the paper.
 In Sec II we work out the equations of the 
propagating sound in a moving fluid using the geometric 
acoustics \cite{LL_hydro:1987}. We consider both relativistic 
and  non-relativistic cases. Then in Sec III we construct the 
Mach cones fomed by sound waves emitted by the associated 
jet, and circles fomed by sound waves emitted by a jet which is 
originated at the same hard collision point but is different from 
the trigger and associated jets. Meanwhile, all the cones and 
circles are solutions of the relativistic and non-relativistic 
equations. In Sec IV we proceed to discussion of the issue 
related to intersection of the Mach cone surface with the fireball's 
freezeout surface in 3D. Finally, in Sec V we calculate the spectra 
of secondaries comparing them with the experiment as well. We 
will use our assumption that most of the contribution to the 
spectra come from the region which is the ``edge'' formed by 
the intersection of the surfaces.

\subsection{Comments on the jet quenching}
  This complicated phenomenon, suggested by Bjorken in 1982, has very long  history,
  which we of course would not go into here. 
 For the purpose of this work it is not  important to know
the  microscopic mechanism of  jet quenching, and for modelling we will
use simple forms of the energy  $dE/dx$ deposited
into ambient matter. Thus this subsection will only include comments
relevant for what follows below.

Hard collision events of partons are a particular type of ``initial state perturbation".
It is important  \cite{Shuryak:2007fu}  that those generates $four$ (not two!)
 jets. Two of them approximately balance
their large transverse momenta 
One of these two jets is the parent of hard ``trigger" hadron, the other  
called the associate jet,  propagates through the medium
and deposits certain amount of energy into the ambient 
matter. Below 
we will follow what happens with it, provided it can be 
described hydrodynamically.

Two more bremstrahlung cones, or jets, are directed forward and backward in the beam direction.
Since these ones do not have large transverse momenta, they are rarely discussed, if at all.
However, they do produce extra multiplicity, roughly the same as two other jets,
which should be seen in the correlation functions. As suggested in \cite{Shuryak:2007fu}, 
those would naturally explain the ``hard ridge", an extra structure near the triger
direction in the azimuthal angle, but long-range in the rapidity. In \cite{I} the main puzzle has been discussed: 
   why
this perturbation remains correlated with the trigger direction rather than create a sound circle
and move $\pm 1$ radian away, creating ``shoulders" instead? The suggested answer was flux tube formation,
natural if hydrodynamics is elevated to dual magnetohydrodynamics.   
In this paper we will however not discuss that option, restraining to the usual sound waves.

  For modelling energy loss one needs to know how the jet energy loss depends on the matter
(entropy) density $s(x)$ as well as the length $L$ travelled by the 
jet in the matter  from the moment of its production. 

The simplest assumptions, resembling what happens in QED, is that it is proportional to matter density
and is independent of x, $-dE/dx\sim s*const(x)$. It did not however took long to notice
 \cite{Shuryak:2001me,Drees:2003zh} that this assumption is in conflict with the 
experimental data,  on jet quenching dependence on the azimuthal angle $\phi$ relative to the impact parameter direction. 
Those ususally are  characterized by the parameter
\be %
v_{2}(p_{t})= <cos(2\phi)>\,,
\ee %
 (not to be confused with the
elliptic flow: we now hard transverse momenta  \,$p_{t} >6 \,GeV$\,). This quantity can also be seen as related to the
difference between the hard hadron spectra from jets travelling in
the $x$ direction ($\phi=0$, in-the-reaction-plane) and in the $y$
direction ($\phi=90^{0}$, out-of-plain). Not going into the
history of the $v_{2}$ debate, let us comment on its current
status. There are two ideas on the market, reproducing the
$v_{2}(p_{t}>6\, GeV)$  data: \\
(i) One \cite{Liao:2008dk}
is that the quenching is not proportional to density but is 
enhanced in the near-$T_c$ region or the
so called ``mixed phase". \\ (ii) Another is  based on the strong coupling theory
(AdS/CFT) picture of matter, resulting in the energy loss  \cite{Marquet:2009,Mueller:2008}
\be %
-{dE \over dx}|_{strong \, coupling} \sim T^{4}x^{2}\,,
\ee %
While this regime may look similar to the weak coupling BDMPS result
 \be -{dE \over dx}|_{weak \, coupling} \sim
T^{3}x \ee 
it is in fact qualitatively different. To see this one should recall that $T\sim 1/x^{1/3}$ due to longitudinal expansion.
As a result, the strong coupling result diverges at large x, while the weak coupling one is spread over the whole path.
Thus the former one does describe the $v_2$ data, 
while the latter one does not.

Two proposed explanations are very different in nature, the near-$T_c$ ``blackness" of matter \cite{Liao:2008dk}
versus physical growth of the jet ``falling into the IR" in \cite{Marquet:2009,Mueller:2008}. 
Yet it is hard to tell them apart in practice, because at RHIC  the near--$T_c$
region corresponds to the proper time $\tau=5-9 \, fm/c$ which is very close to the typical time it takes for the
associate  jet to traverse the fireball. Perhaps one would need to do a detailed comparison of RHIC and LHC results to see which one describes the data better, as at LHC the near--$T_c$ region is
shifted to later times.
 
 Another feature of the strong coupling results is that large fraction of energy is released at the stopping point, see details
in \cite{Chesler:2008uy}. In
our calculations below, there is one more reason for the importance of the endpoint
of the jet path inside the matter:  the less time perturbation in matter has to travel,
the larger sound amplitude it has.

\section{Sounds in a moving fluid and the geometric acoustics }

As detailed in hydrodynamics textbooks (e.g., by Landau-Lifshitz
\cite{LL_hydro:1987}) the rays describing the sound propagation
can be described in the ``geometric acoustics" approximation
which uses the analogy between the Hamilton-Jacobi equation for
the particle and the wave sound equation.
 The velocity and
direction of the sound propagation can be derived with the help of
the eikonal function. 
The  Hamilton equations of motion for
``phonons" (or ``particles'' of the sound) have  the following generic
form
\ba %
{d\vec{r} \over dt}= {\partial \omega(\vec{k},\vec{r}) \over \partial \vec{k}}  \label{eqn_speed}\,,  \\\
{d\vec{k} \over dt}=-{\partial \omega(\vec{k},\vec{r}) \over
\partial \vec{r}}\,, 
\label{eqn_force}
\ea %
driven by the (position dependent) dispersion relation
$\omega(\vec{k},\vec{r})$. For clarity, let us start with the
simplest non-relativistic case, namely of small velocity of
the flow, $u \ll 1$. In this case the dispersion relation is
obtained from that in the fluid at rest by a  Galilean
transformation, so that
\be %
\omega(\vec{k},\vec{r})=c_{s} k + \vec{k}\vec{u}\,.
\label{eqn_nonrel}
\ee %
In the trivial case of the constant flow vector, $\vec
u=const(r)$, the first of the above equations just obtains an additive
correction 
\be %
{d\vec{r} \over dt} = c_{s} \vec{n}_{\vec{k}} +\vec{u}\,,
\ee %
where $\vec{n}_{\vec{k}}=\vec{k}/k$ is the unit vector in the
direction of the momentum. The second equation remains trivial
\be %
{d\vec{k} \over dt} = 0\,,
\ee %
as there is no coordinate dependence anywhere. So,  a point perturbation
in a moving fluid simply produces a moving sound circle.
motion with the speed of sound, then the position

After this ``warm-up", let us do a non-trivial problem,
 simple enough   to be analytically solved. Let us consider a (generalized) Hubble-like
flow
\be %
u_{i}(r)=H_{ij} r_{j}\,,
\label{eqn_u}
\ee %
with some time and coordinate independent Hubble tensor $H_{ij} $. The equation
(\ref{eqn_force}) now reads
\be %
{dk_{i} \over dt} = - H_{ij} k_{j}\,.
\ee %
So, if the Hubble tensor is symmetric and can be diagonalized with
real eigenvalues $H_{1},H_{2},H_{3}$, then the solution in its
eigenframe is exponential momentum contraction
$k_{i}(t)=exp(-H_{i}t) k_{i}(0)$.  However, if the
Hubble tensor contains an anti-symmetric part, the eigenvalues can be imaginary, so that the $\vec{k}$ is
 rotating around the vector $\epsilon_{ijk}H_{jk}$.

The first equation (\ref{eqn_speed})  reads :
\be %
{dr_{i} \over dt} = c_{s} \vec{n}_{\vec{k}}(t) + H_{ij}
r_{j}(t)\,.
\ee %
In the simplest case when the Hubble matrix is proportional to
the unit matrix  the
solution is simply a linear addition of the Hubble expansion and
sound motion:
\be %
\vec{r}(t)= c_{s} t \vec{n}_{\vec{k}} + \vec{r}(0)\exp(+Ht)\,.
\ee %

  This example can be applied to real hydro explosion, in which the late-time flow can indeed be well approximated by a Hubble form.
  For central collisions we will only discuss, the Hubble tensor is even isotropic. The only complications is that this pattern needs some time to
  be developed, so below we will use it only later than some time $t_{0} \approx 5\,fm$ \cite{hydro}. So the 
 two-dimensional form of  the equations
(\ref{eqn_speed}) and (\ref{eqn_force})  to be used is
\ba %
{d k_{1} \over dt} & = & - H k_{1}(t)\,\Theta(t - t_{0})
\nonumber\\
{d k_{2} \over dt} & = & - H k_{2}(t)\,\Theta(t - t_{0})
\nonumber\\
\nonumber\\
\nonumber\\
{d r_{1} \over dt} & = & c_{s}\cos{(\theta)} + H r_{1}(t)\,\Theta(t
- t_{0})
\nonumber\\
{d r_{2} \over dt} & = & c_{s} \sin {(\theta)}
 + H r_{2}(t)\,\Theta(t - t_{0})\,. 
\label{eqn_system1}
\ea %
 The initial conditions
to this system are specified as follows:
\ba %
k_{1}(0) & = & \cos{(\theta)}\,,\,\,\,\,k_{2}(0) = \sin{(\theta)}\,,
\nonumber\\
r_{1}(0) & = & x(0)\,,\,\,\,\,r_{2}(0) = y(0)\,, 
\label{eqn_initial}
\ea %
where $x(0)$ and $y(0)$ are constants and measured by $fm$\,'s.
They are the coordinates of the hard collision point where a jet
is produced and simultaneously emits phonons.
For the angle $\theta$ we take the value of the Mach angle
relative to the jet velocity from equation (\ref{eqn_Mach}):
$\cos{(\theta)}\equiv\cos{(\theta_{M})}=<c_{s}>/v_{jet}$.

The corresponding analytical solutions will be
\ba %
k_{1}(t) & = & \cos{(\theta)}\,e^{-Ht\,\Theta(t - t_{0})}
\nonumber\\
k_{2}(t) & = & \sin{(\theta)}\,e^{-Ht\,\Theta(t -  t_{0})}
\nonumber\\
r_{1}(t) & = & \left[ c_{s}\cos{(\theta)}\,t + x(0)
\right]\Theta(t -  t_{0}) +
\nonumber\\
&+& \left[ {c_{s}\cos{(\theta)}\left( 1 - e^{H(t- t_{0})} -
5H\,e^{H(t - t_{0})} \right) \over H} - \right.
\nonumber\\
& &\left. - x(0)\,e^{H(t - t_{0})} \right](\Theta(t -  t_{0}) - 1)
\nonumber\\ 
r_{2}(t) & = & \left[ c_{s}\sin{(\theta)}\,t + y(0)
\right]\Theta(t -  t_{0}) +
\nonumber\\
&+& \left[ {c_{s}\sin{(\theta)}\left( 1 - e^{H(t - t_{0})} -
5H\,e^{H(t -  t_{0})} \right) \over H} - \right.
\nonumber\\
& &\left. - y(0)\,e^{H(t -  t_{0})} \right](\Theta(t -  t_{0}) - 1)\,.
\label{eqn_solution}
\ea %

A bunch of rays emitted to all directions from some initial point
will thus be at a surface of an expanding sphere, as in constant
flow:
\be %
(\vec{r}(t) -\vec{r}(0) \exp(+ Ht) )^{2} = t^{2} c_{s}^{2}\,,
\ee %
with a sphere center moving ``with a flow".

In heavy ion collisions the explosion is relativistic, with
transverse velocity reaching $u\sim 0.7c$. Therefore the local
Galilean transformation should be substituted by the Lorentz one,
which gives
\be %
\omega(\vec{k},\vec{r})= c_{s} k \,\cosh{\!(Y)} + \vec{k}\hat{\vec{u}}
\,\sinh{\!(Y)} \label{eqn_rel}
\ee %
where $\hat{u}$ means a unit vector in the direction of $u$, and
$Y$ is the rapidity. Now it is no longer possible to separate the
equations (\ref{eqn_speed}) and (\ref{eqn_force}), and they should be
solved together numerically. Substitution of the equation
(\ref{eqn_rel}) into (\ref{eqn_speed}) and (\ref{eqn_force}) gives
us the following system of four differential (relativistic)
equations:

\begin{figure}[t!]
  \vskip 0.3in
  \hskip -0.2in
  \includegraphics[width=5cm]{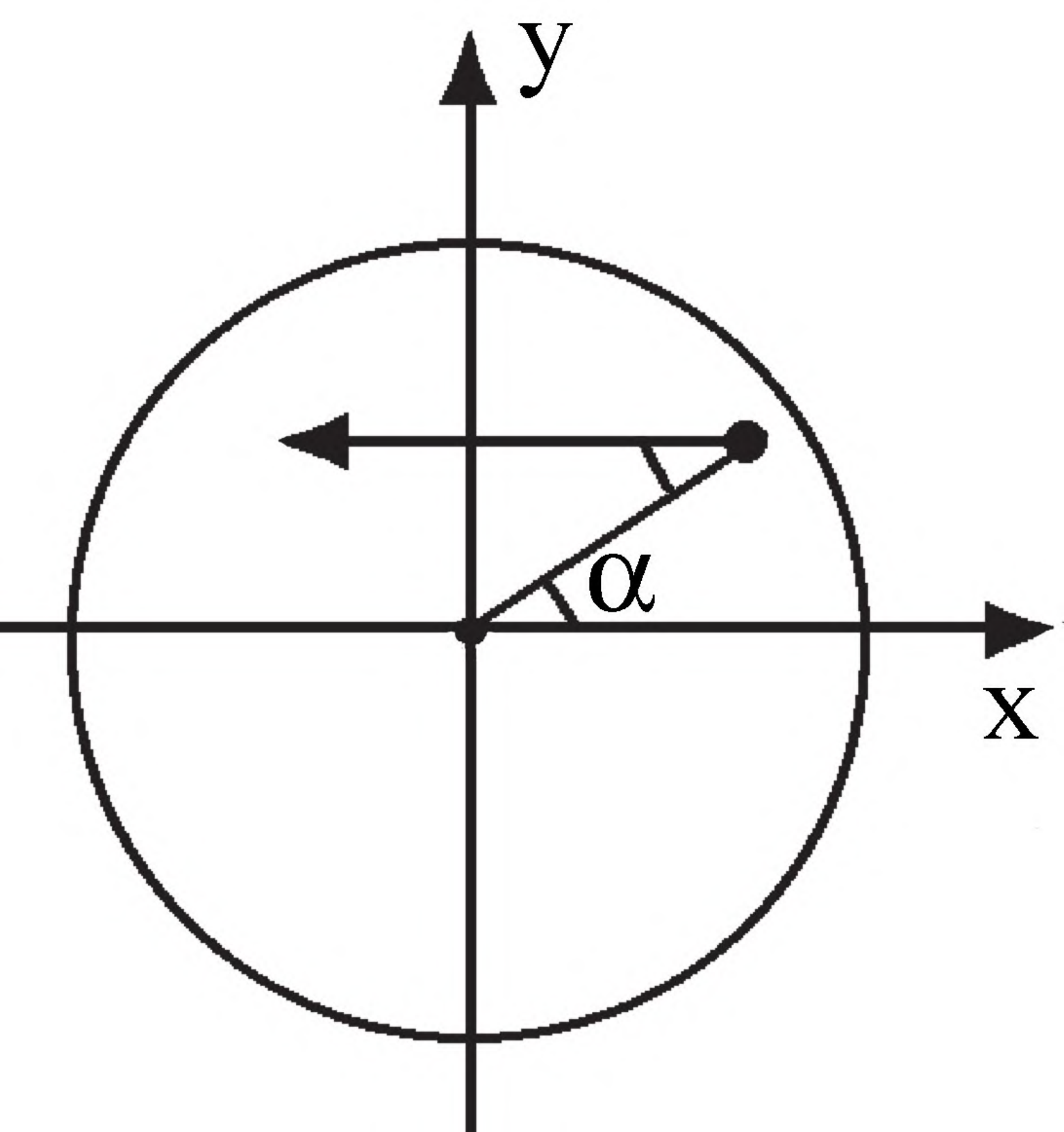}
  \caption{\label{fig_alpha} The $\alpha$ is the angle between the 
   associated jet's momentum and the axis $x$.  The arrow inside the circle is the momentum
   of the associated jet which originates from the hard collision point opposite
   to the trigger jet (not shown) in the x direction.}
\end{figure}

\begin{figure*}[ht!]
  \vskip 0.2in
  \hskip -0.2in
  \includegraphics[width=8.cm]{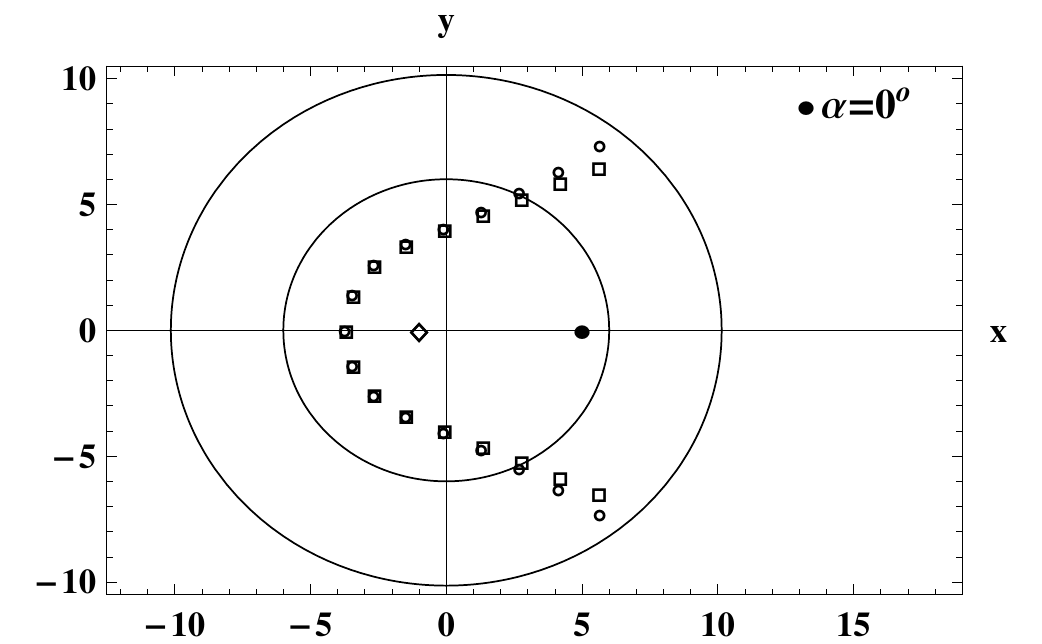}
    \includegraphics[width=8.cm]{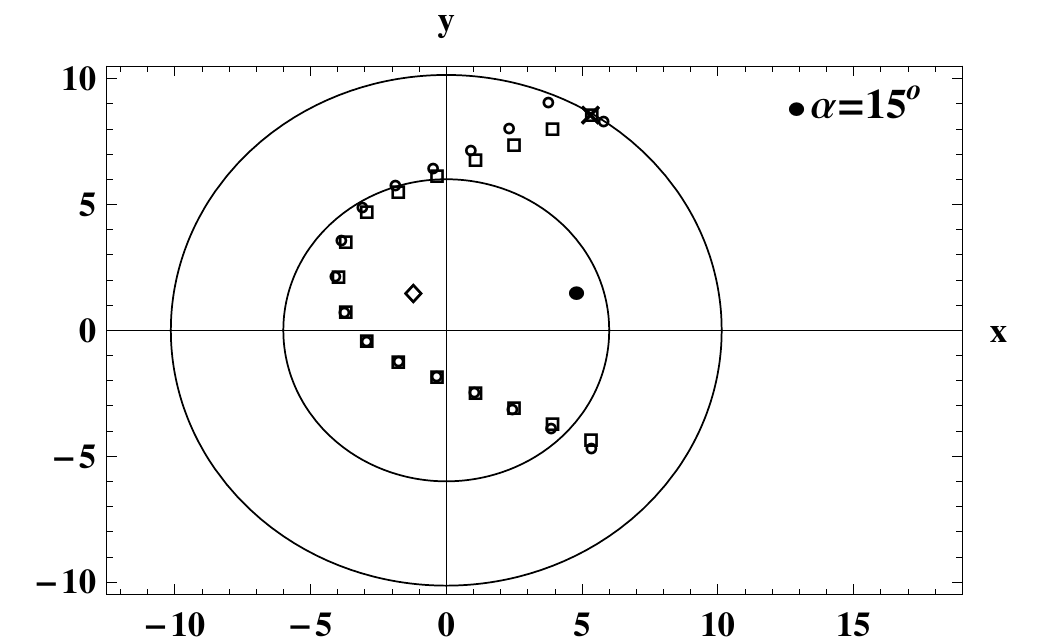}
     \includegraphics[width=8.cm]{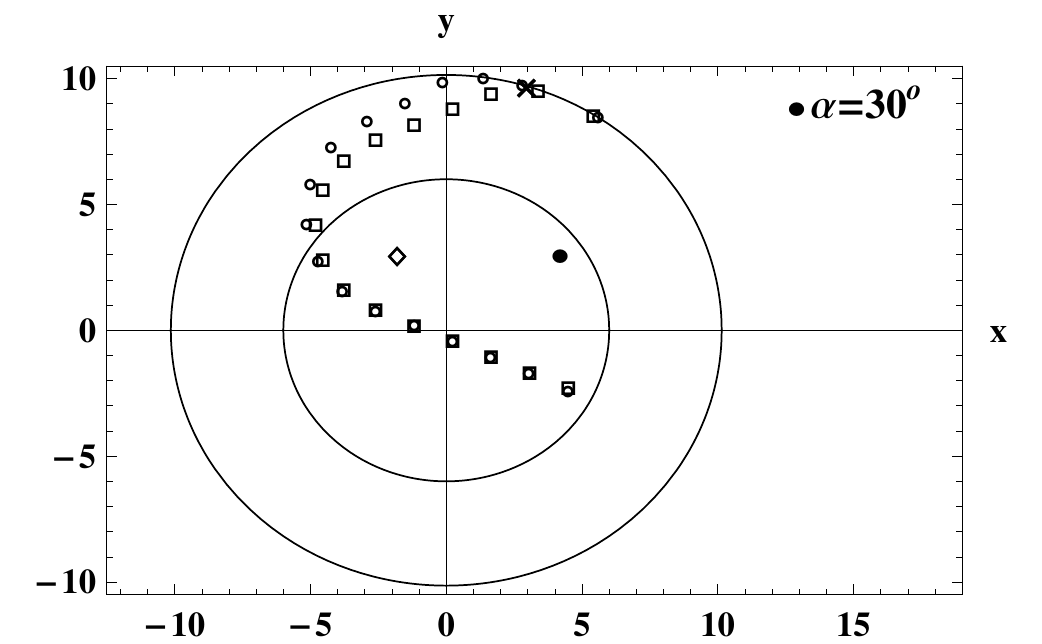} 
     \includegraphics[width=8.cm]{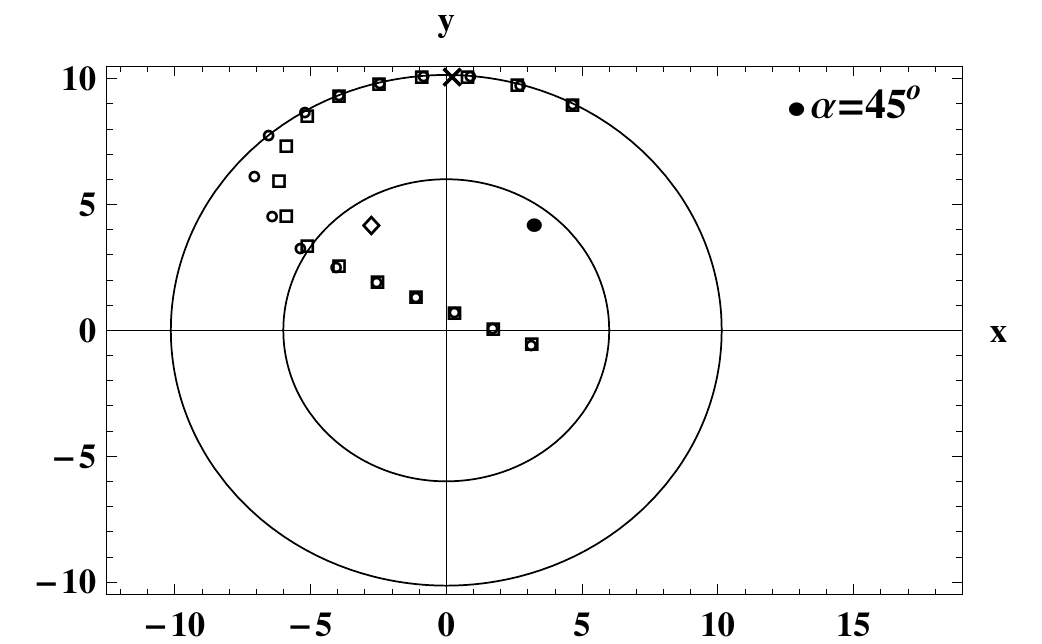}
  \includegraphics[width=8.cm]{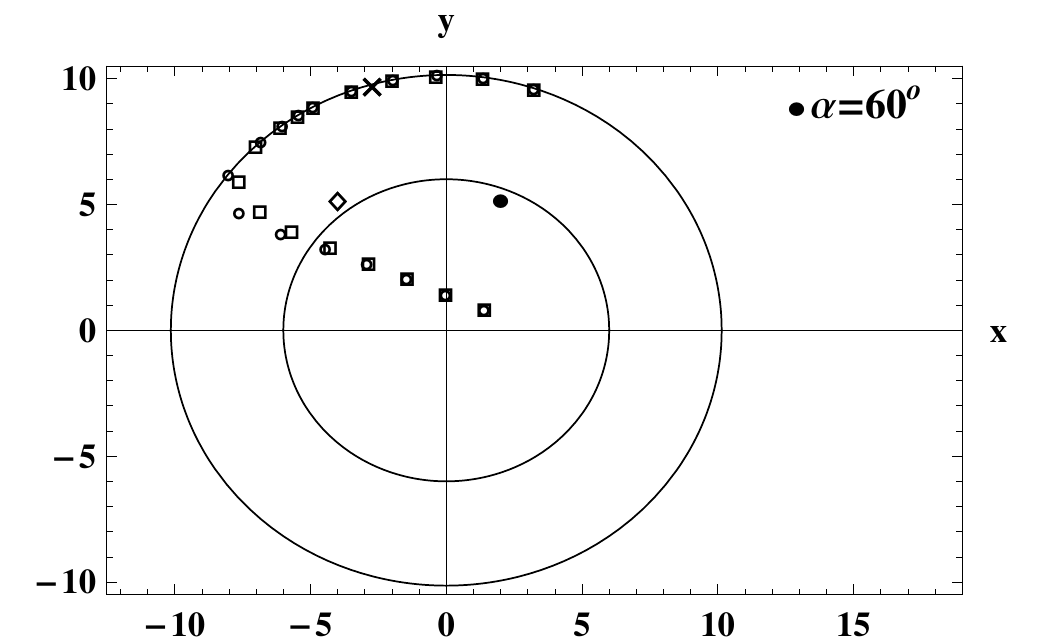}
  \includegraphics[width=8.cm]{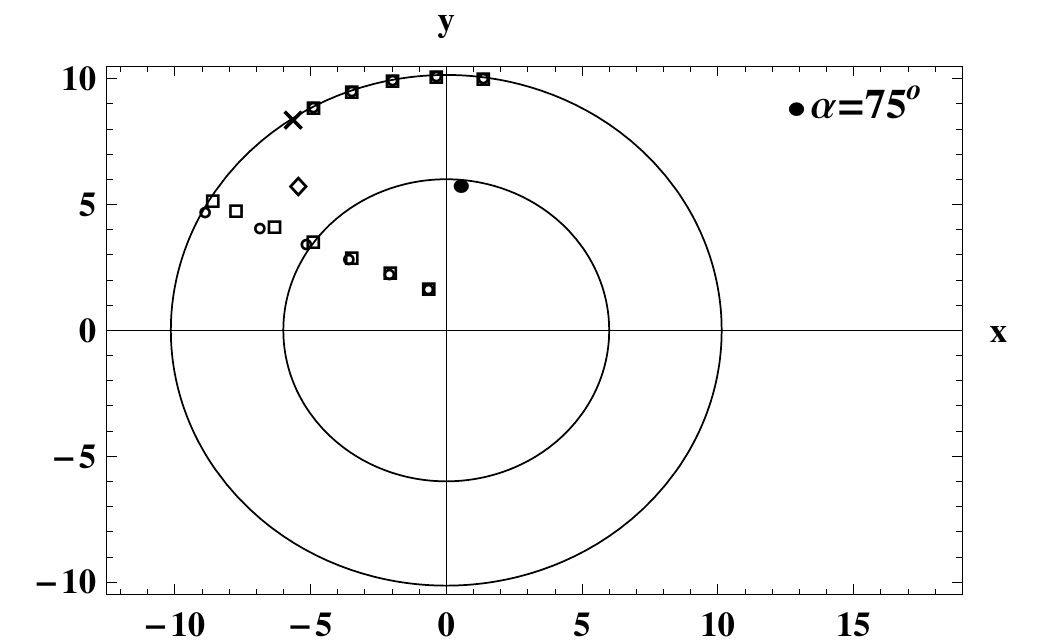}
  \caption{\label{fig_Cone_1}(Color online) A Mach cone originated from a jet moving to the left from the 
  point $x(0) = 5.0\,fm$ and $y(0) = 0\,fm$ at the angle $\alpha = 0$. The path of the jet corresponds to 
  \,$L_{stopping} = 6\,fm$,\, the radius of the inner circle - $R_{in} = 6\,fm$, and the radius of the outer 
  circle, containing the fireball at its highest extension \,$R_{f} = \exp\!{(H\!\cdot\!(t_{f} - t_{0}))} \simeq 10.14\,fm$.\, The cone created by the open
  squares is the solution of the non-relativistic equations (\ref{eqn_system1}), and the cone created by the 
  open circles is the solution of the relativistic equations (\ref{eqn_system2}). The filled circle shows the hard
  collision point, and the open diamond shows the stopping point of the jet. The values of the angle $\alpha$ shown in each figure, is defined in Fig.1.}
\end{figure*}

\ba %
{d k_{1} \over dt} & = & - c_{s} \sqrt{k_{1}^{\,2} +
k_{2}^{\,2}}\,\times
\nonumber\\
& \times & {\partial \over \partial r_{1}}\!\left( 1 \over
\sqrt{1 - H^{2}(r_{1}^{\,2} + r_{2}^{\,2})\Theta(t -  t_{0})}\right) -
\nonumber\\
& - & k_{1}\,\,{\partial \over
\partial r_{1}}\!\left( H r_{1}\Theta(t -
 t_{0}) \over \sqrt{1 - H^{2}(r_{1}^{\,2} + r_{2}^{\,2})\Theta(t -  t_{0})}
\right) -
\nonumber\\
& - & k_{2}\,\,{\partial \over
\partial r_{1}}\!\left( H r_{2}\Theta(t -
 t_{0}) \over \sqrt{1 - H^{2}(r_{1}^{\,2} + r_{2}^{\,2})\Theta(t -  t_{0})}
\right)
\nonumber\\
\nonumber\\
{d k_{2} \over dt} & = & - c_{s} \sqrt{k_{1}^{\,2} +
k_{2}^{\,2}}\,\times
\nonumber\\
& \times & {\partial \over \partial r_{2}}\!\left( 1 \over
\sqrt{1 - H^{2}(r_{1}^{\,2} + r_{2}^{\,2})\Theta(t -  t_{0})} \right) -
\nonumber\\
& - & k_{1}\,\,{\partial \over
\partial r_{2}}\!\left( H r_{1}\Theta(t -
 t_{0}) \over \sqrt{1 - H^{2}(r_{1}^{\,2} + r_{2}^{\,2})\Theta(t - t_{0})}
\right) -
\nonumber\\
& - & k_{2}\,\,{\partial \over
\partial r_{2}}\!\left( H r_{2}\Theta(t -
 t_{0}) \over \sqrt{1 - H^{2}(r_{1}^{\,2} + r_{2}^{\,2})\Theta(t -  t_{0})}
\right)
\nonumber\\
\nonumber\\
{d r_{1} \over dt} & = & c_{s} {k_{1} \over \sqrt{k_{1}^{\,2} +
k_{2}^{\,2}}} {1 \over \sqrt{1 - H^{2}(r_{1}^{\,2} +
r_{2}^{\,2})\Theta(t -  t_{0})}} +
\nonumber\\
& + & {H r_{1}\Theta(t -  t_{0}) \over \sqrt{1 - H^{2}(r_{1}^{\,2} +
r_{2}^{\,2})\Theta(t -  t_{0})}}
\nonumber\\
\nonumber\\
\nonumber\\
{d r_{2} \over dt} & = & c_{s} {k_{2} \over \sqrt{k_{1}^{\,2} +
k_{2}^{\,2}}} {1 \over \sqrt{1 - H^{2}(r_{1}^{\,2} +
r_{2}^{\,2})\Theta(t -  t_{0})}} +
\nonumber\\
& + & {H r_{2}\Theta(t -  t_{0}) \over \sqrt{1 - H^{2}(r_{1}^{\,2} +
r_{2}^{\,2})\Theta(t -  t_{0})}}\,, \label{eqn_system2}
\ea %
with the same initial conditions as in (\ref{eqn_initial}). 

We solve the systems of equations (\ref{eqn_system1}) and 
(\ref{eqn_system2}) at freezeout time \,$t_{f}=12\,fm$,\, 
speed of sound \,$c_{s} = 0.4$,\, and the Hubble constant
\,$H = 0.075\,fm^{-1}$.\, The $t_{f}$  is the time at which 
the fireball reaches the freeze-out temperature. (In Appendix A,
we clarify  a value for the freezout time used, along with 
establishing the kinetic freezeout temperature.)

\begin{figure*}[t]
  \includegraphics[width=8cm]{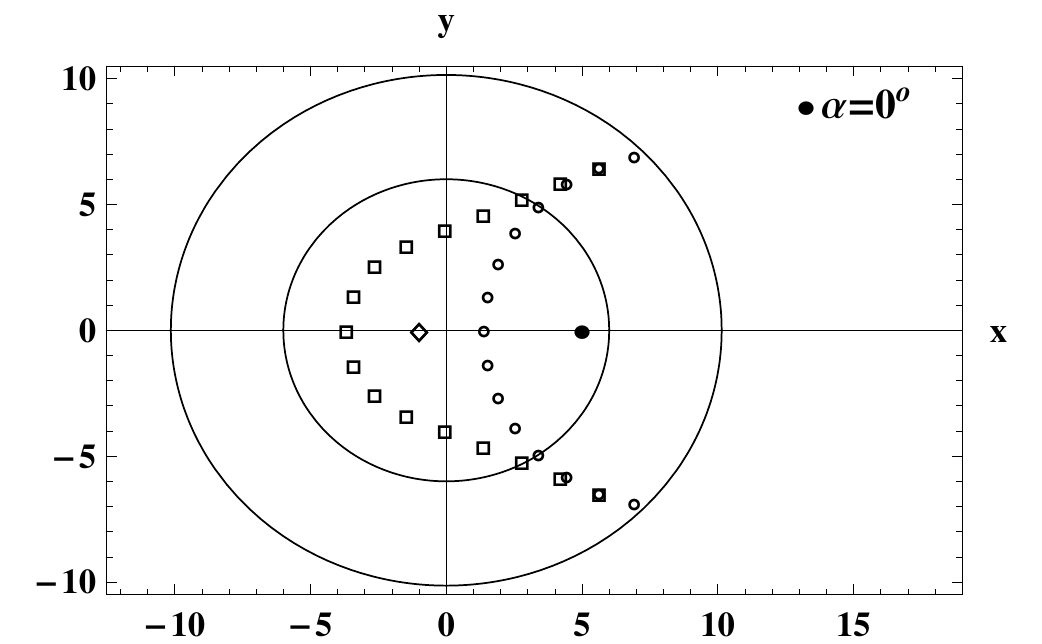} 
    \includegraphics[width=8cm]{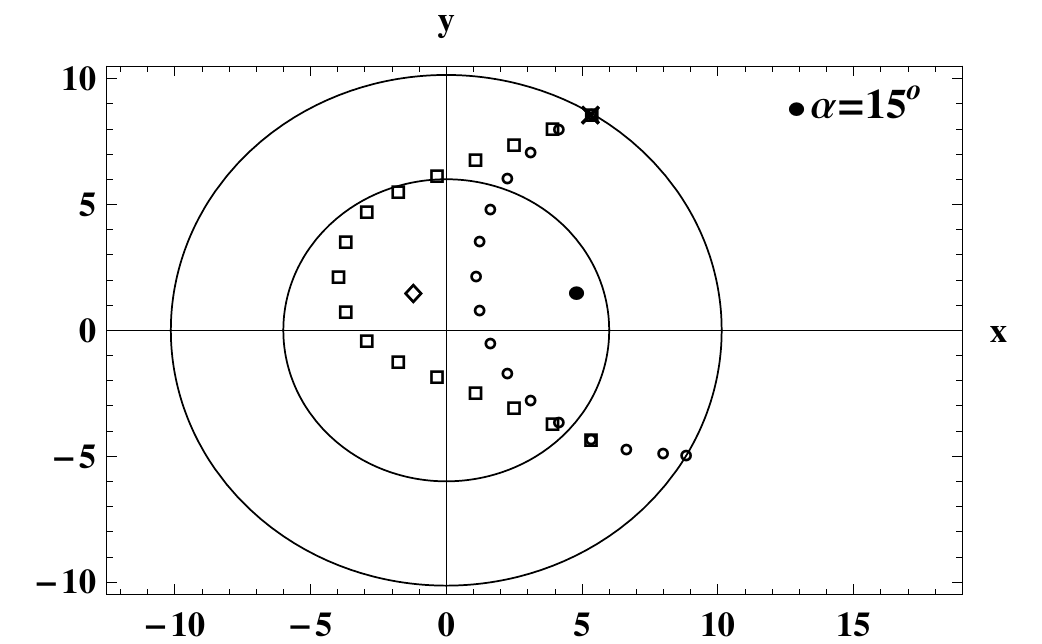}
      \includegraphics[width=8cm]{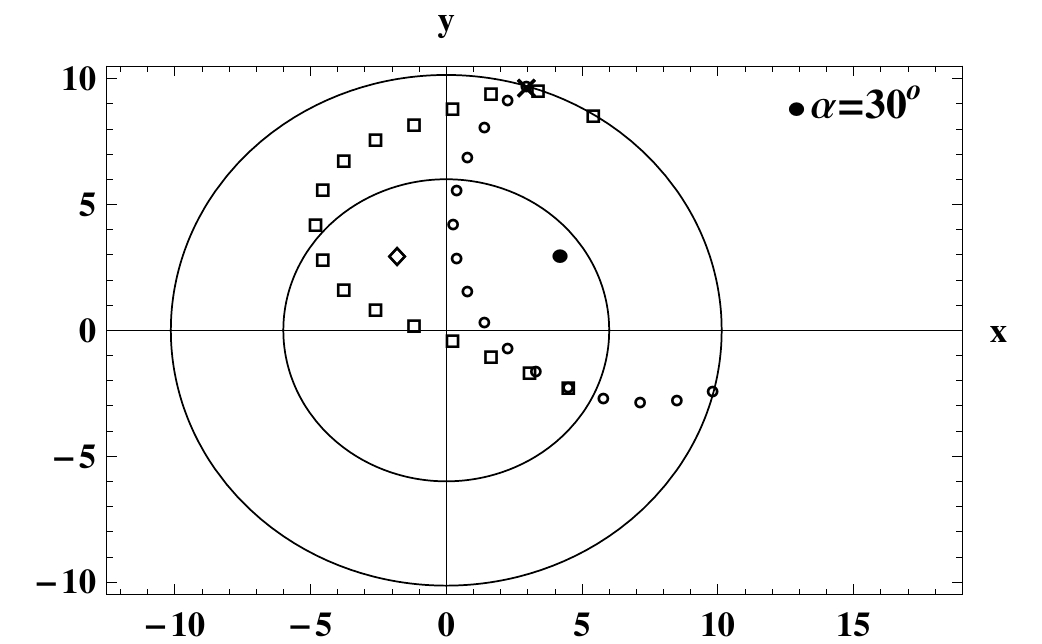}
        \includegraphics[width=8cm]{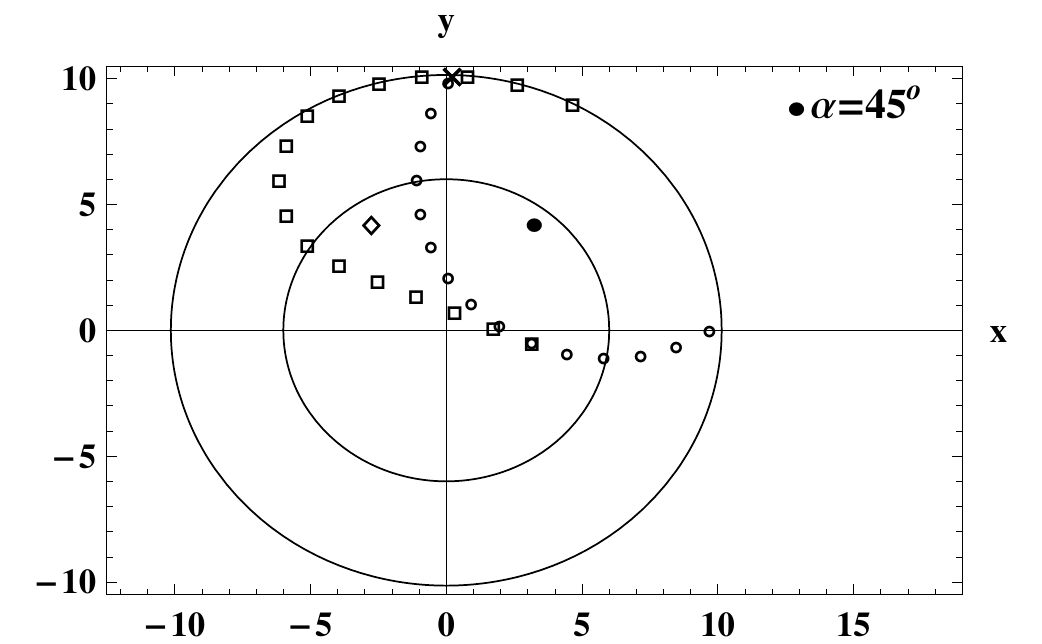}
          \includegraphics[width=8cm]{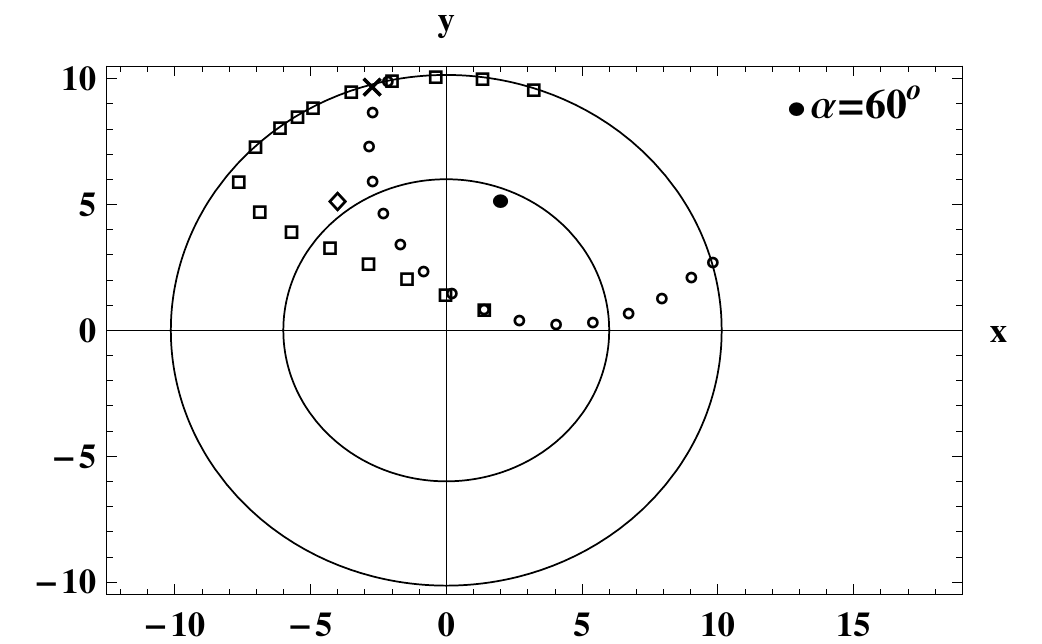}
            \includegraphics[width=8cm]{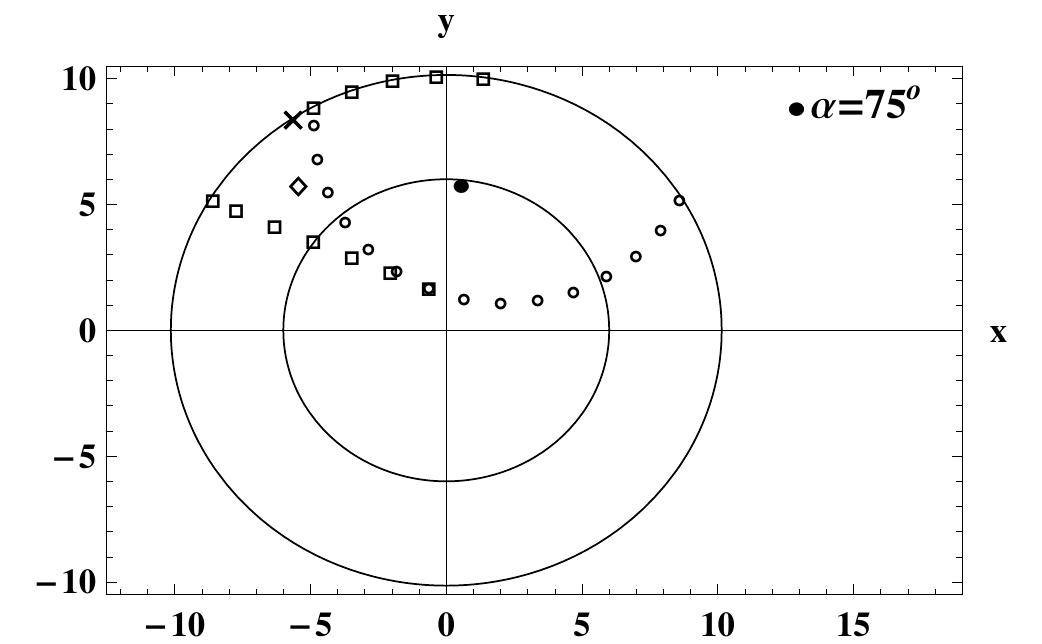}
  \caption{\label{fig_Cone_Ring_1}  The Mach cone (open squares) as in Fig.\,\ref{fig_Cone_1}, and  
   the ring (open circles) which is originated from the hard collision point $x(0) =5.0\,fm$ and $y(0) =0\,fm$ at the angle    
   $\alpha = 0$. These two structures are the solutions of the non-relativistic equations (\ref{eqn_system1}). }
\end{figure*}

\section{Distorted cones and circles of the expanding fireball}

In order to simplify  geometry of the collision, we consider only
 central $Au-Au$ collisions. The notations are explained in Fig.\,\ref{fig_alpha}: jets which have different impact parameters in respect to the fireball
 are defined via the  
angle $\alpha$, between the direction of the associate jets' momentum and the 
axis $x$ in which the trigger jets go. 
(We do not show the trigger jet in Fig.\,\ref{fig_alpha}) as a reminder that we do not include
the sounds from it in what follows, considering its energy loss  to be negligible. Due to trigger bias effect,
we place hard scattering events at fixed distance, 1 fm, from the nuclear edge.)

Quenching of the associate jet is the source of the phonons, emitted from its path along the Mach direction.
After that they move along the phonon rays as detailed in the previous section. We show only  solutions in the 
upper part of the transverse plane of the fireball, binning angle $\alpha$ into six bins 
 $\alpha=0$, $15^{0}$, $30^{0}$, $45^{0}$, $60^{0}$ 
and $75^{0}$, which are depicted in Fig.\,\ref{fig_Cone_1}, 
 The case of $\alpha=90^{0}$ 
does not give rise to any conical structure since the jet at this point 
only strikes the edge of the fireball/inner sphere leaving it totally, 
without making phonons propagate in the medium. The solutions in 
the lower part are symmetrical to those in the upper part, 
that's why they are not shown.
The conical and ring structures  are the 
solutions of the systems of equations: non-relativistc (\ref{eqn_system1}) 
and relativistic (\ref{eqn_system2}).  

 Note that in all these six figures the 
jets move to the left with the speed of light, from points of their 
production, located inside the inner circle at a distance $1\,fm$ from 
that circle which in fact is the surface of a nucleus. This inner circle 
corresponds to the case of \,$t_{f}=0\,fm$,\, when the 
firebal/inner sphere initially is not expanded as a whole. The outer 
circle corresponds to the radius \,$t_{f}=12\,fm$,\,it shows
 the maximal extension of the fireball reached by this time. 

The jet quenching discussed in those figures assume the stopping 
distance of the jets be \,$L_{stopping} = 6\,fm$, shown by the diamonds. ( 
For comparison purposes we have also calculated cones for the stopping 
distance \,$L_{stopping} = 12\,fm$\, not shown, with similar 
results to those  in Figs.\,\ref{fig_Cone_1} except for different ``rounding"
of it does not matter much.)
This means that we assume the jet energy and the energy loss are related by
$E/|dE/dx|= 6\, fm$. The actual values of $E,dE/dx$ are not  important for these plots and will
matter only later, when we will determine the actual spectra modification.   

We see that at the angles increasing towards higher $\alpha$, 
the conical structure created by the phonons has a 
distorted/crooked shape, which is due to the Hubble expansion 
starting from $t_{0}\approx 5\,fm$. The jets, produced at low and 
intermediate angles, do not reach the edge of the outer sphere, 
meaning that they do not leave the fireball. However, at the 
higher angles, some jets go out from the fireball at some values 
of the time within the range of \,$t_{f}=5-12\,fm$. One conclusion 
is that in the fireball with the given kinematical and dynamical 
conditions, the shapes of the Mach cones do not differ from each 
other significantly obtained from the non-relativistic equations 
(\ref{eqn_system1}), and from relativistic equations 
(\ref{eqn_system2}). Therefore, hereinafter we can consider only 
one of these cases, namely one can take the non-relativistic case 
for carring out the other calculations.

 We already mentioned above  two more jets which are also created by a
hard collision which  move perpendicularly  
to the transverse plane along the beam directions. 
These two longitudinal jets are approximately rapidity-independent source which also may emit 
the sound waves.  Their propagation we  calculated 
using the same equations (\ref{eqn_system1}) but with the initial conditions
 in (\ref{eqn_initial}), where $\theta$ ranges from $0$ to $360^{0}$. 
Such (rapidity-independent) sound circles  are depicted in Figs.\,\ref{fig_Cone_Ring_1} together with the cones
(which are for one rapidity only, namely that of the associate jet). We will not use those circles below: in fact we have calculated them
in order to compare to what is calculated in \cite{III} by a different method.

\section{Intersection of the Mach surface with timelike and spacelike freezeout surfaces}

Standard  expression for a particle spectrum, known as Cooper-Frye formula 
\cite{Cooper:1974mv}, is a thermal spectrum boosted by the 4-vector of the flow velocity $u^{\mu}$ and 
integrated over the $3-dimensional$ freezeout surface. 
\be %
 E{dN\over d^{3}p} = \int_{\sigma_{t} + \sigma_{s}} d\Sigma_{\mu} p^{\mu}\,
f\!\!\left( {p^{\nu} u_{\nu} \over T}\right)\,,
\label{eqn_Cooper}
\ee %
where $\sigma_{t}$ and $\sigma_{s}$ are time-like and space-like parts of the freezeout surface,
respectively. The function $f$ is Bose/Fermi/Boltzmann distribution, for pions it is
\be %
f\!\!\left( {p^{\nu} u_{\nu} \over T}\right) = {1 \over \exp(-p^{\nu} u_{\nu}/T)\pm 1} 
\label{eqn_function}
\ee %
simply corresponds to the thermal distribution of matter inside the fluid cells. (If needed, the function can be appended by
an anisotropic part corresponding to viscosity.)

\begin{figure}[t]
\begin{center}
  \includegraphics*[width=8cm]{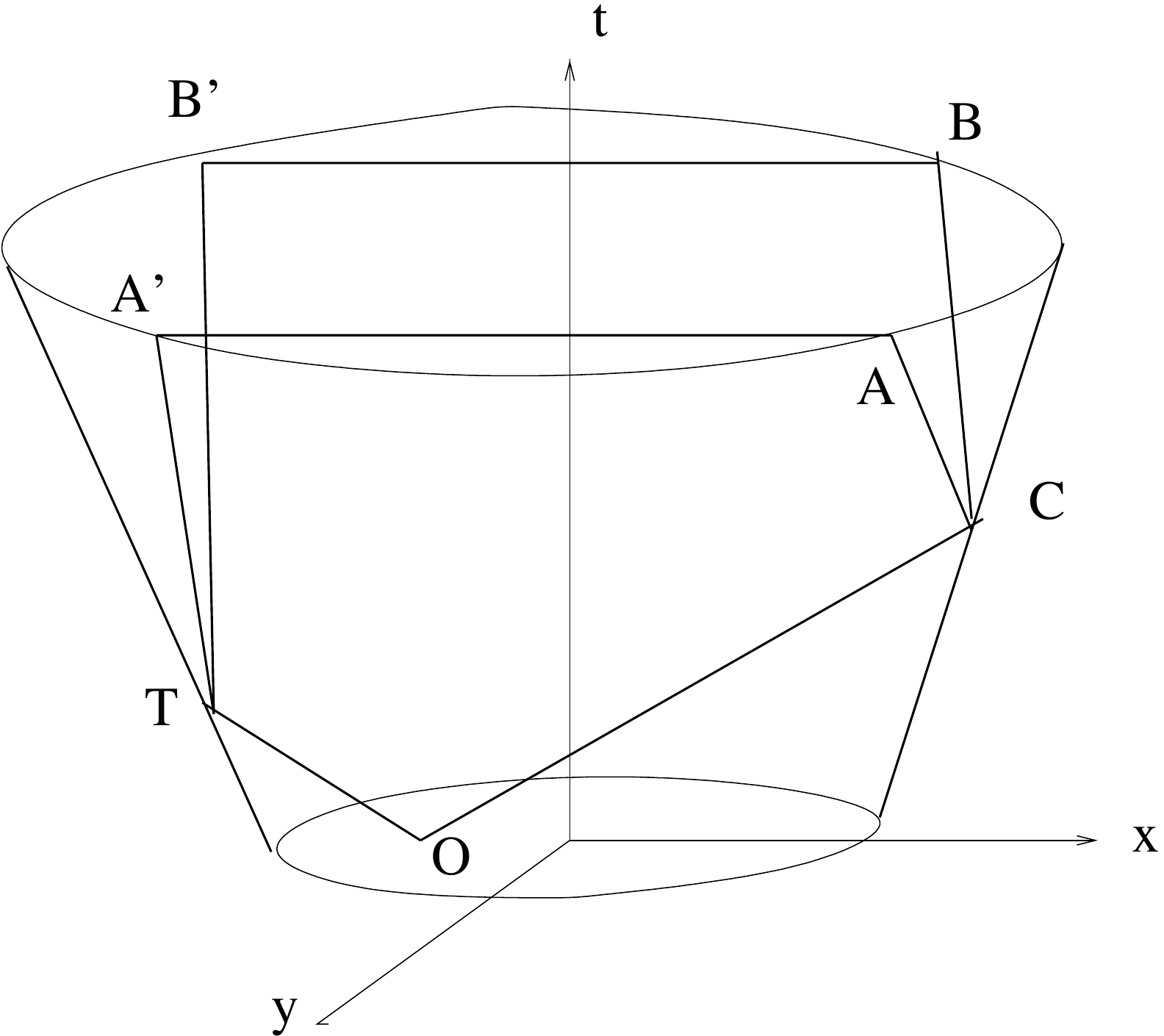}
  \vskip -0.1in
  \hskip 0.5in
  \caption{\label{fig_cone_at_tf} The schematic shape of the Mach surface in the transverse
   $(x,y)$ plane at \,$z = 0$\, and fixed time (upper plot), as well as its shape in $3d$ including the 
   proper longitudinal time (lower plot). The Mach surface $\sigma_{M}$ is made of two parts, $OCAA'T$ 
   and $OCBB'T$. This figure is from \cite{Shuryak:2011}.}
\end{center}
\end{figure}

  We will be interested in a special case of this distribution  in the region where, at one hand, the
hydrodynamics works well, and on the other, the ratio $(p_{t}/T_{f}$ is as large as possible.
(As we are only interested in the tail of the function $f$, so that the Boltzmann 
approximation will always be enough.)
This ratio, if large, strongly enhances the effect of the small perturbation, the sounds, on the particle spectrum.
In fact, for
\,$p_{t}=1\,GeV$\,  and \,$p_{t}=2\,GeV$\ we will be using below, this large ratio time small amplitudes of the sound waves 
is able to produce effects of the order O(1) in the exponent. This interplay
of large $p_t/T_f$ and small sound wave is one of the core ideas of this paper.

  Before we go to details about realistic freezeout surfaces, in the next section and Appendix A, 
 we would like to introduce another qualitative idea, this time related with the radial flow.   
The concept has been introduced in 
 \cite{Shuryak:2011} and is explained in schematic way in 
Fig.\,\ref{fig_cone_at_tf}.  
(Note that in Fig.\,\ref{fig_cone_at_tf} it is assumed that the 
associated jet propagates  to the right, unlike in the conventions above.) 
It shows the picture 
in $3d$,  at fixed longitudinal coordinate \,$z = 0$. 
The proper time $\tau$ runs upward, 
and the upper circle schematically represents the time-like part of
the freezeout surface, $\sigma_{t}$, approximated by the constant
time surface, \,$t = t_{f}$.\, The lower circle is the so called 
``initiation time surface'', and the conical surface connecting the
two ellipses is our approximation to the space-like part of the 
freezeout surface, $\sigma_{s}$. The points $A$, $B$, $A'$ and 
$B'$  are the intersection of the sound waves with 
 $\sigma_{t}$, and the region between them contains matter affected by the waves: outside it
 is not affected, by hydro causality. The points
 $T,C$ are intersections of the trigger and the companion jets with $\sigma_s$.  
The main idea is that the vicinity of four points $A$, $B$, $A'$ and 
$B'$ should dominate the spectra (modified by the sounds) because the large upper circle
is where the radial flow is at its maximum. (We will return to this idea at the quantative level in the next section.)




  Let us now promote this figure from 3-dimension to all 4, by
adding the longitudinal $z$ direction.  We cannot plot it in full, and therefore limit ourselves to the $t=t_f$ time slice.
Since we have the intercept of 3 surfaces (Mach,$\sigma_t,\sigma_s$) in 4d, it must be a 1-dimensional curve. It is in fact
 two ellipses, one having points $A$, $B$ on it,
and the other having $A'$, $B'$. Let us call these two edges 
$\epsilon_{C}$ and $\epsilon_{T}$, for  the companion (associated) 
jet and the trigger jet, respectively. (Since the trigger-bias forces the 
associated jet to deposit much larger amount of energy, the 
former one has much larger chance to become visible.)

\begin{figure}[h!]
  \vskip 0.0in
  \hskip -0.2in
  \includegraphics[width=9cm]{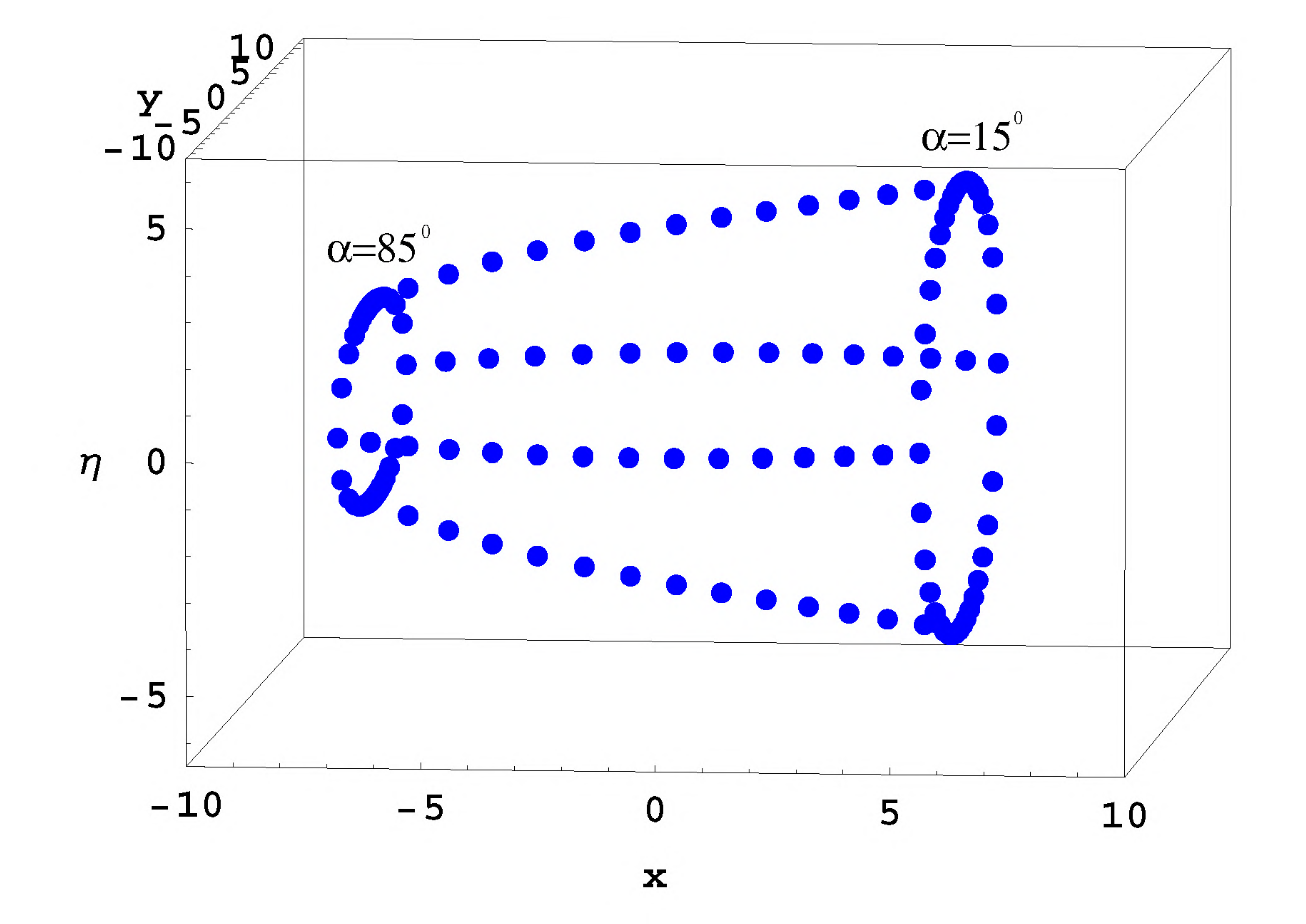}
  \caption{\label{fig_Ellipse_1}(Color online) Two elliptic structures correspond to the edges $\epsilon_{C}$
   produced by phonons at intersection of all the three surfaces. These phonons in turn are emitted by an 
   associated jet which moves in the $x$ direction (or parallel to it) emitting the phonons in $x$, $y$ and 
  $\eta$ directions. The largest ellipse corresponds to the jet with \,$\alpha = 15^{0}$,\, and the smallest ellpise  
   corresponds to the jet with \,$\alpha = 85^{0}$.}
\end{figure}

\begin{figure}[h!]
  \vskip 0.3in
  \hskip -0.2in
  \includegraphics[width=9cm]{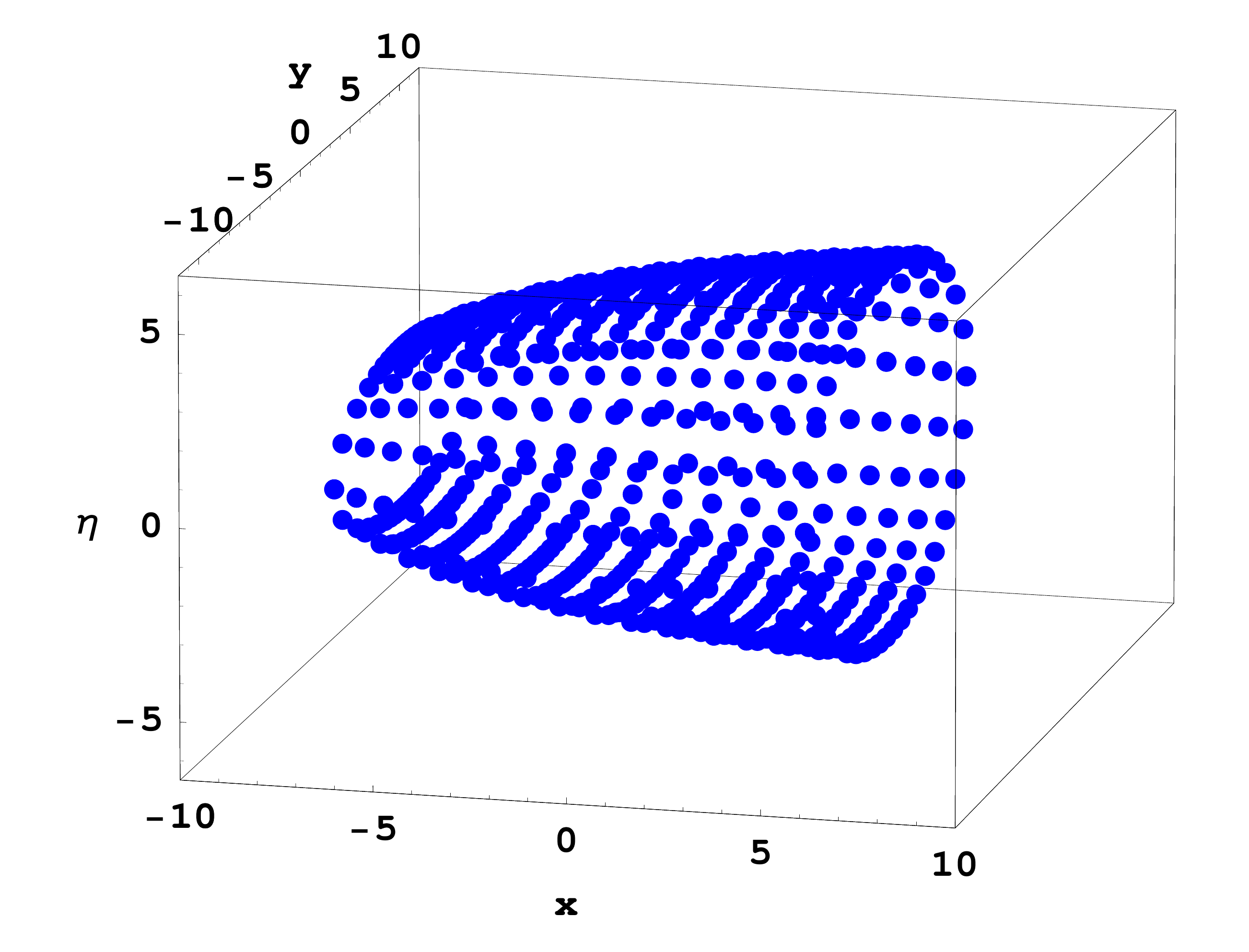}
  \caption{\label{fig_Ellipse_2}(Color online) The same figure as in Fig.\,\ref{fig_Ellipse_1} but which is complete
   including the contributions of all the jets with \,$\alpha = 15^{0} \div 85^{0}$.}
\end{figure}

\begin{figure}[h!]
  \vskip 0.0in
  \hskip -0.2in
  \includegraphics[width=9cm]{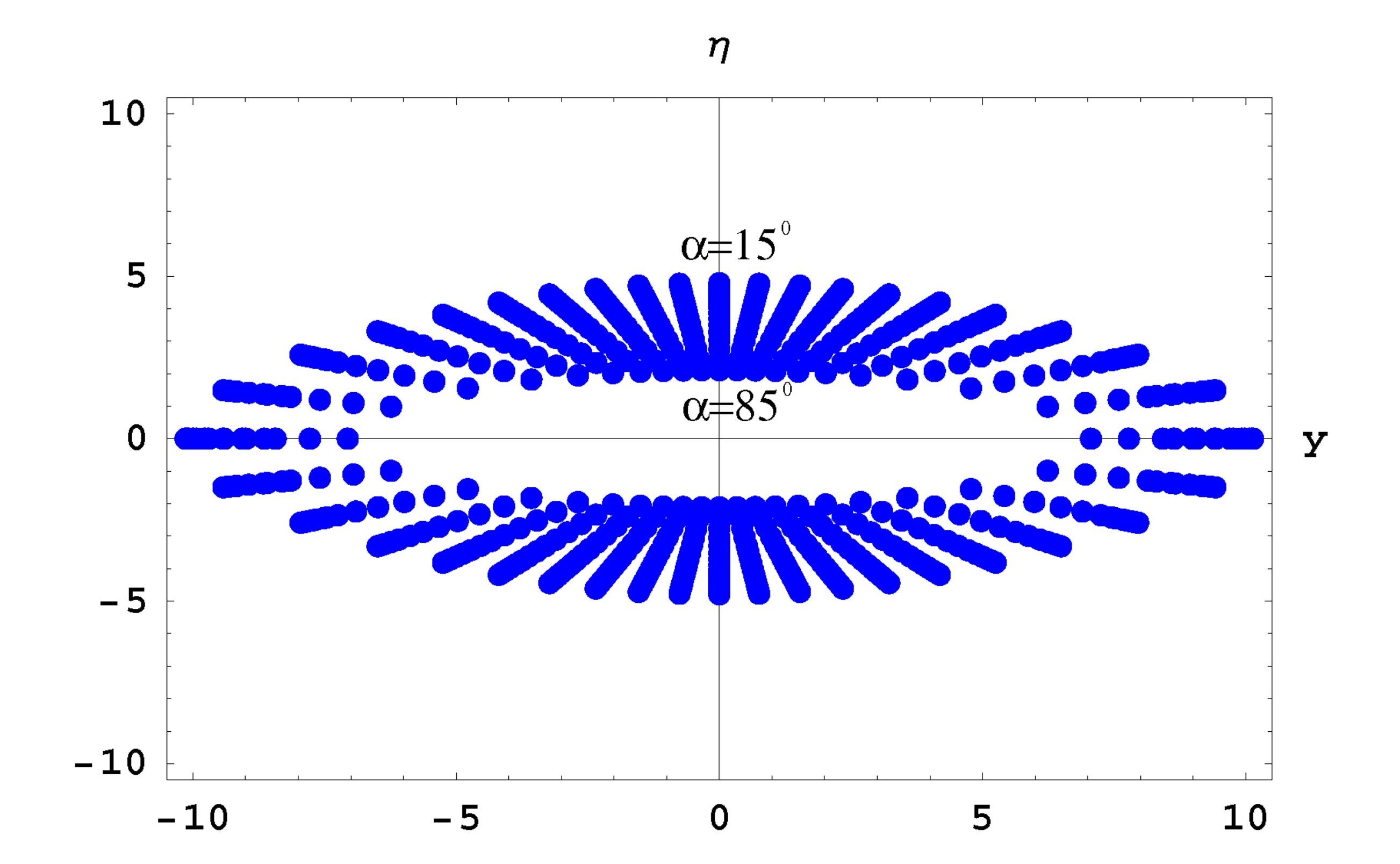}
  \caption{\label{fig_Ellipse_3}(Color online) The same figure as in Fig.\,\ref{fig_Ellipse_2} but in the $(\eta,y)$ plane.
   The largest ellipse corresponds to the jet with \,$\alpha = 15^{0}$,\, and the smallest ellpise corresponds to the jet 
   with \,$\alpha = 85^{0}$.}
\end{figure}

\begin{figure}[h!]
  \vskip 0.3in
  \hskip -0.2in
  \includegraphics[width=9cm]{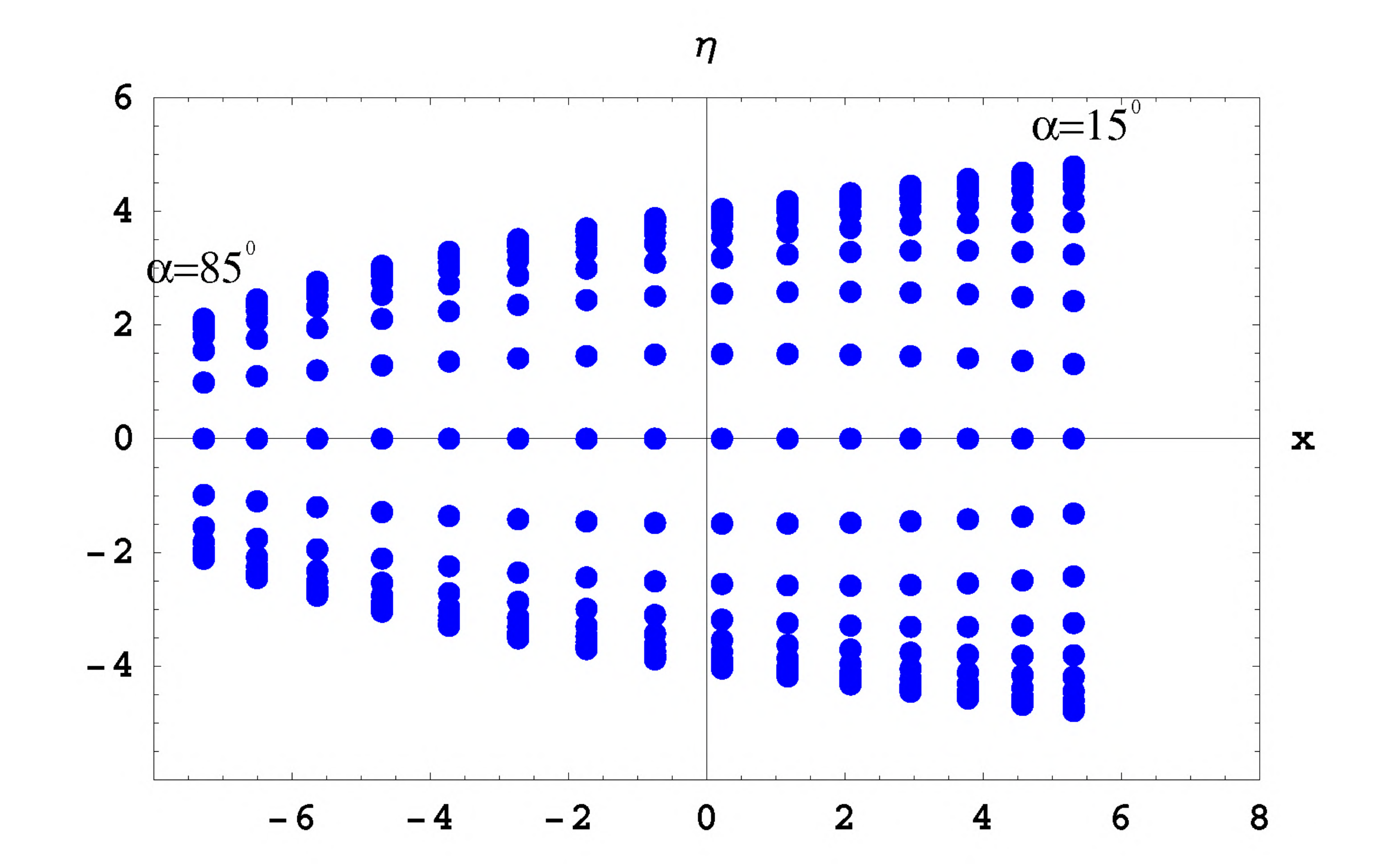}
  \caption{\label{fig_Ellipse_4}(Color online) The same figure as in Fig.\,\ref{fig_Ellipse_2} but in the $(\eta,x)$ plane.
   The largest vertical corresponds to the jet with \,$\alpha = 15^{0}$,\, and the smallest vertical corresponds to the jet  
   with \,$\alpha = 85^{0}$.}
\end{figure}

Let us calculate the extension of these elliptic curves into the longitudinal coordinate
 $z$ direction, or rather its proxi,  the 
pseudo-rapidity $\eta$.  In Figs.\,\ref{fig_Ellipse_1},  
\ref{fig_Ellipse_2}, \ref{fig_Ellipse_3} and \ref{fig_Ellipse_4}
we show correspondingly the edges $\epsilon_{C}$ 
(ellipses) created by phonons at \,$t_{f} = 12\,fm$,\, emitted 
by a jet whose momentum is directed as in Figs.\,\ref{fig_Cone_1}, 
 However, now 
the increase of the angle $\alpha$ from the previous jet
to the next one is $\Delta\alpha = 5^{0}$, rather than $15^{0}$. 
Again the case of the jet produced at \,$\alpha = 90^{0}$\, 
is ignored. Besides, we also consider phonons (from all the jets under 
consideration) moving in the lower hemisphere of the fireball.
The total number of all the considered jets (events) is thirty.

\section{Application to particle spectra with the modified freezeout surface}
The temperature and velocity in Cooper-Fry formula  have complicated space-time 
dependence derived from hydrodynamics. As in \cite{III} we assume that the 
freezeout surface is an isotherm of the form of \,$T(t,r)=T_{f}$.\, Accordingly, 
the time-like part of the  surface, for example, can be written as
\be
\Sigma^{\mu} = (t_{f}(x,y),x,y,\eta)\,.
\ee
So at the time \,$t_{f}(x,y)$\, the fireball reaches the freezeout temperature $T_{f}$.
Note that (\ref{eqn_Cooper}) contains the vector normal to the surface.
As to the element of the surface, it is represented as
\ba
d\Sigma_{\mu} & = & \sqrt{-g}\epsilon_{\mu \nu \lambda
\rho}\frac{\partial \Sigma^{\nu}}{\partial x} \frac{\partial
\Sigma^{\lambda}}{\partial y} \frac{\partial
\Sigma^{\rho}}{\partial \eta} dxdyd\eta  =
\nonumber\\
& = & \left(-1,\frac{\partial t_{f}}{\partial
x},\frac{\partial t_{f}}{\partial y},0\right)t_{f}
dxdyd\eta
\nonumber\\
& = & - t_{f}dxdyd\eta + t_{f}dt_{f}dyd\eta + t_{f}dt_{f}dxd\eta\,,
\label{eqn_surface}
\ea
where $g$ is the determinant of the metric and $\epsilon_{\mu \nu
\lambda \rho}$ is the Levi-Civita symbol.

For $p_{\mu}d\Sigma_{\mu}$ and  $p_{\mu}u^{\mu}$ one may use the formulae 
from  Ref.\,\cite{BaRo:2007} which are the following:
\ba
p_{\mu}d\Sigma^{\mu} & = & \left( m_{t}\cosh(\eta - y) - { \over } \right.
\nonumber\\
& - & \left.  p_{t}\cos(\phi - \phi_{p}){dt_{f}(r) \over dr} \right)\!\!t_{f}(r)\,rdrd\phi d\eta ,
\label{eqn_Cooper1}
\ea
\be %
p_{\mu}u^{\mu} & = & \left( m_{t}\cosh(\eta - y)u^{\tau} 
-  p_{t}\cos(\phi - \phi_{p})u^{r} \right) ,
\label{eqn_Cooper2}
\ee
where \,$u^{\tau} = \gamma = (1- u(r)^{2})^{-1/2}$,\,  \,$u^{r} = u(r) u^{\tau}$\, 
(with$ u(r)$ given by  (\ref{eqn_u})), \,$m_{t} = \sqrt{p_{t}^{2} + m_{\pi}^2}$\, 
(the $m_{\pi}$ is the pion mass). We extract the function $t_{f}(r)$ from  hydro solutions discussed 
in Appendix A.

\begin{figure}[h!]
  \vskip 0.3in
  \hskip -0.2in
  \includegraphics[width=8cm]{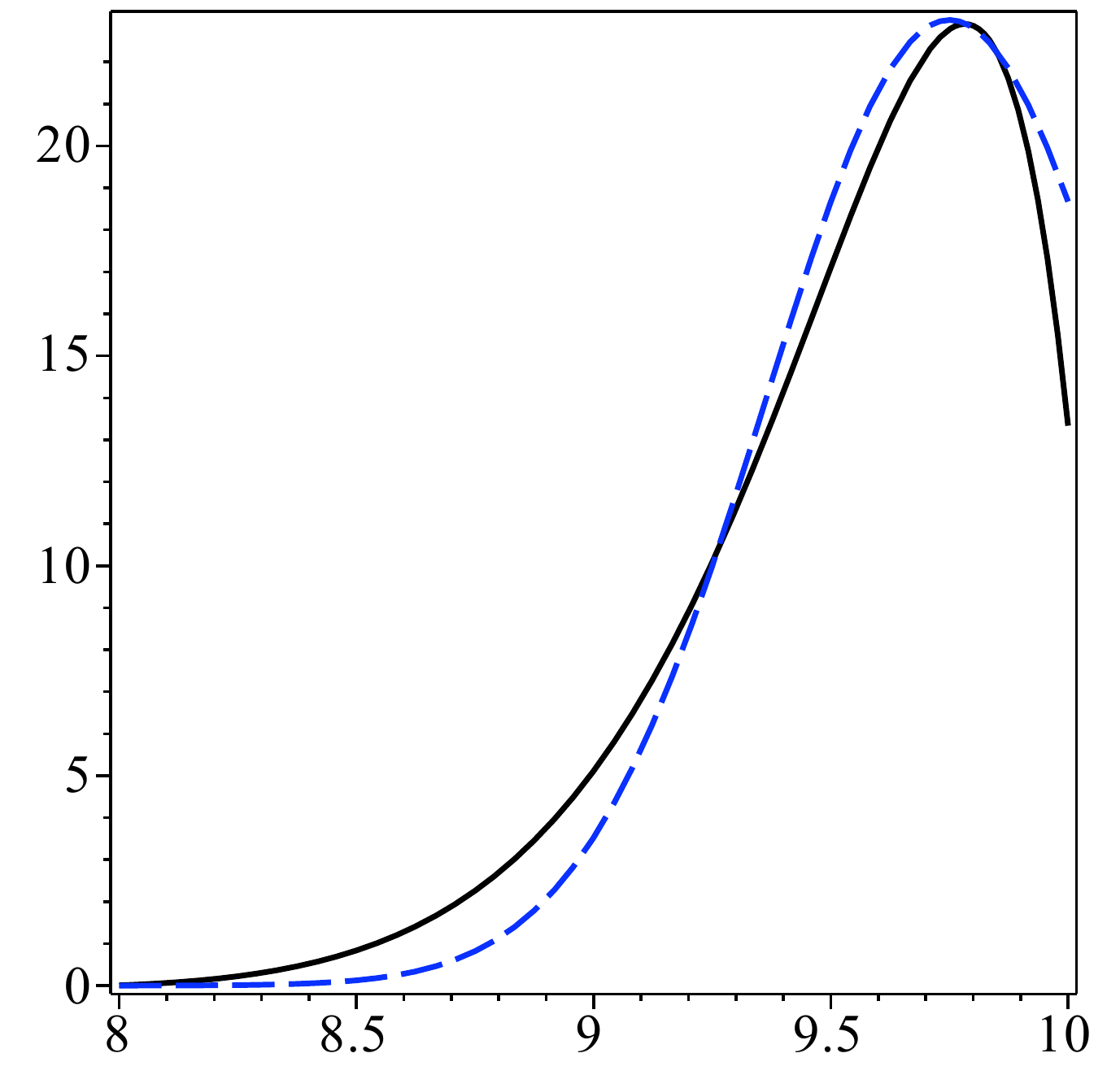}
 \caption{\label{fig_peak}(Color online)  The contribution to the pion spectrum
 (arbitrary units)  vs the radial coordinate $r$ (fm). The solid (black) curve is 
 the Cooper-Fry formula for $p_t=2 \, GeV,T_f=.12\, GeV$, the dash (blue) curve is the Gaussian approximation mentioned in the text. }
 \end{figure}

For each $p_t$ and the direction of the particle $\phi_p$ there is a point on the freezeout
surface which maximally contribute to the spectrum. Indeed, the radial flow grows with
$r$ and enhance the spectrum, but the surface itself makes a turn and ends.
One can determine the spectrum from the Cooper-Fry formula and compare it to
e.g. Gaussian approximation around this point. Those two are shown in Fig. \ref{fig_peak}, as a particular example.
The lesson is that there exist a sharp peak near the fireball's edge, at which the quest for the strongest radial flow and for
the largest amount of surface area (and thus multiplicity) are in the best compromise. We have checked that those peaks
indeed produce the spectra in agreement with what hydro papers reported, and with what are observed experimentally.
This confirms that the tails of the spectra we are interested are dominated by the vicinity of such points.

Now we  consider the effect of sound perturbations on the particles 
(pions) spectra. Here we follow the discussion of \cite{III} since the 
same process occurs in our case.
The perturbations can affect the spectra in two ways. We have two
variables which have their contributions: namely, the flow velocity
$u_{\mu}$ and temperature $T$.  The first effect is related to the $u_{\mu}$  
in the exponent of (\ref{eqn_Boltzmann}) which should be corrected 
by an extra term of the first order due to sound. The second effect is 
related to the first order $T$ perturbations. One can consider the so called 
``zeroth'' order fireball and the fireball with the ``hot spot'' induced by an 
initial-state perturbation. In the presence of this ``hot spot'' the corresponding 
perturbation in temperature, $\delta T$, is positive and implies a production 
of extra particles as compared to the zeroth order fireball. The $\delta T$ 
locally delays the freezeout temperature by the formula
\be %
T_0(t,r) + \delta T(t,r) = T_{f} \,,
\ee %
which also provides an extra volume for containing the produced
extra particles. Thus the freezeout surface is deformed in terms of
the temperature and volume. Generally, it provides a little larger flow
as compared to that of the zeroth order fireball.

The velocity perturbation, $\delta u_{\mu}$, added to the velocity $u_{\mu}$ gives
\be
u_{\mu} \rightarrow u_{\mu} + \delta u_{\mu}\,.
\ee

In order to find $\delta T$ and $\delta u$ we make use of well known
hydrodynamical relations:
\be %
n = {g' \over \pi^{2}\hbar^{3}}T^{3}\,,
\label{eqn_hydro1}
\ee
\be %
\epsilon = 3p = 3nT\,.
\label{eqn_hydro2}
\ee
The $g'$ is the total degeneracy factor which counts the total number of 
degrees of freedom, summed over the spins, flavors, charge (particle-antiparticle) 
and colors of particles. For the QGP, \,$g' \approx 3$ at freezeout.\, From  the 
equations (\ref{eqn_hydro1}) and  (\ref{eqn_hydro2}) one obtains
\be %
{\delta T} = \left( {\pi^{2}\hbar^{3} \over 12g'} \right){{\delta\epsilon} \over T^{3}}\,.
\label{eqn_temp_energ}
\ee
In the denominator we can take \,$T^{3} \approx T_{f}^{3}$. 

The next step is to use the first order velocity perturbation from  
Ref.\,\cite{LL_hydro:1987} which is of the form of 
\be %
\delta u = - c_{s}{\delta\epsilon \over (\epsilon_{0} + p_{0})}\,.
\label{eqn_vel_energ}
\ee
With the corresponding substitutions from (\ref{eqn_hydro1}), (\ref{eqn_hydro2})
and (\ref{eqn_temp_energ}), the equation (\ref{eqn_vel_energ}) results in
\ba %
\delta\vec{u}^{(1)} & = & -3c_{s}\,{T^{3} \over T_{0}^{4}}\,{\delta T}\,\vec{n}_{\vec{k}}  \approx 
\nonumber\\
& \approx & - 3c_{s}\,{\delta T \over T_{f}}\,\vec{n}_{\vec{k}}\,,
\label{eqn_vel_temp}
\ea %
where $\vec{n}_{\vec{k}}$ is the unit vector in the direction of the momentum 
of phonons which are locared in the points of the ``crosses" in the transverse $xy$ 
plane, such as in Fig.2.
Thus we have the equations  (\ref{eqn_temp_energ}) and (\ref{eqn_vel_temp})
which determine the first order perturbations of the temperature and
velocity. The only unknown is the perturbation in energy density, 
$\delta\epsilon$, which should be found as well. In the appendix we work
out the procedure by which the $\delta\epsilon$ is determined from
the wave equation (see Appendix B for more details).

However, there is also another effect related to the $\delta u$ which is of the 
first order of perturbation due to expansion of the fireball: \,$\delta u = H\delta r$,\,
where $r$ is the radial distance from the center of the fireball. 
From our hydro parameterizations 
we have found an approximate function for the temperature
(at freezeout time $t_{f} = 12\,fm$) as a function of $r$.
 It is approximated as
\be %
\!\!\!\!\!\!\!
T(t_{f},r) \approx \left( {-1.75\,r^{2} \over fm^{2}} + {19.20\,r \over fm} + 99.67 \right)\!\!MeV .
\label{eqn_approx}
\ee %
It is relevant to note that the $\delta r$ gives the increase of the fireball's volume
(also giving rise to $\delta T$).
It is equal to
\be %
\delta r = {\delta T(t,r) \over [{\partial T(t_{f},r)} / {\partial r}]|_{r_{f}}}\,.
\label{eqn_der}
\ee %


The $\delta T$ is determined from (\ref{eqn_temp_energ}) which, e.g., for 
\,$dE/dx = 1\,GeV$\, turns out to be $4 \div 5\,MeV$ depending on the angle $\phi$. 
Then due to the effect of the volume change we will have
\be %
\delta \vec{u}^{(2)} & = & H\delta{r}\,\vec{n}_{\vec{r}} =
\nonumber\\
& =  & H\,{\delta T(t,r) \over [{\partial T(t_{f},r)} / {\partial r}]|_{r_{f}}}\,\vec{n}_{\vec{r}}\,,
\label{eqn_vel_dist}
\ee %
where $\vec{n}_{\vec{r}}$ is the unit vector in the radial direction of the
phonons at the points of the ``crosses" in the transverse $xy$ 
plane, such as in Fig.2.
Thus taking into account all the above-mentioned effects, for the exponent in 
the Cooper-Frye formula one can write
\be %
{p^{\mu} u_{\mu} \over T_{f}} = {p_{0} u_{0} \over T_{f}}  + 
{\vec{p}\,\vec{u}\over T_{f}}\,.
\label{eqn_exp}
\ee %
With the perturbations one obtains
\be %
{\vec{p}\,\vec{u }\over T_{f}} \rightarrow {\vec{p}\,\vec{u}\over T_{f}} + 
{\vec{p}\,\delta\vec{u}^{(1)}\over T_{f}} +  {\vec{p}\,\delta\vec{u}^{(2)}\over T_{f}}\,,
\label{eqn_exp1}
\ee %
where \,$|\delta\vec{u}^{(1)}| = \delta u^{(1)}$\, determined from  (\ref{eqn_vel_temp}),
and \,$|\delta\vec{u}^{(2)}| = \delta u^{(2)}$\, determined from  (\ref{eqn_vel_dist}).

In heavy ion collisions some amount of secondaries originate from 
jets produced via hard scattering. When the transverse momentum 
$p_t$ is large enough, tha hard component of it (in the spectrum of 
particles) ``suppresses'' the hydrodinamical spectrum, though the jet 
quenching can be significant. So the hydrodynamics does not work 
at large $p_t$'s. Such a case occurs also at relatively large viscous 
corrections to the viscous term in the stress tensor. These corrections 
get larger with increase of $p_{t}$, and will not be small at some point 
when one compares them with the ideal term of the stress tensor.

\section{Comparison to the RHIC data}

\begin{figure}[t!]
  \vskip 0.3in
  \hskip -0.2in
  \includegraphics[width=7cm]{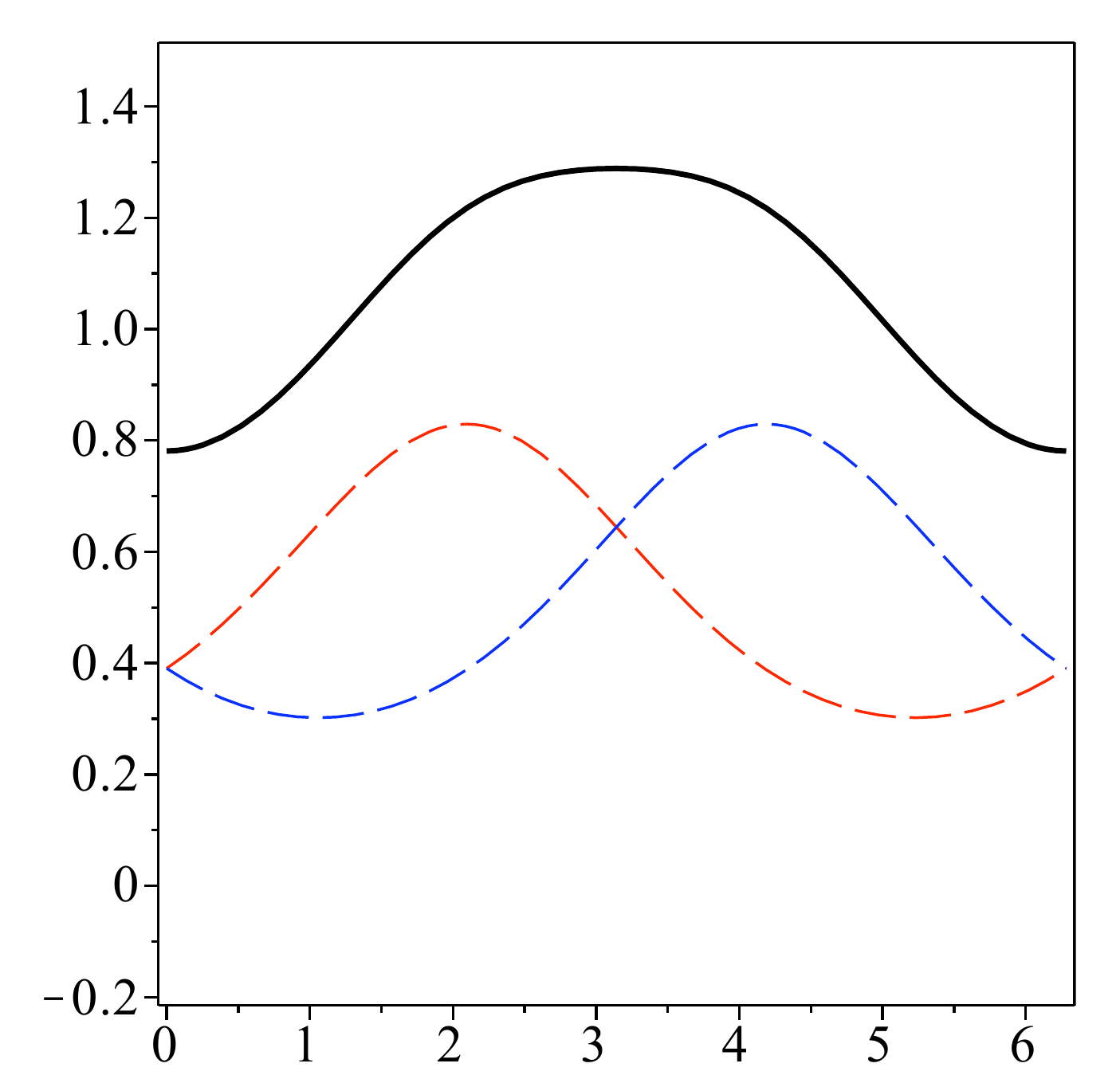}
    \includegraphics[width=7cm]{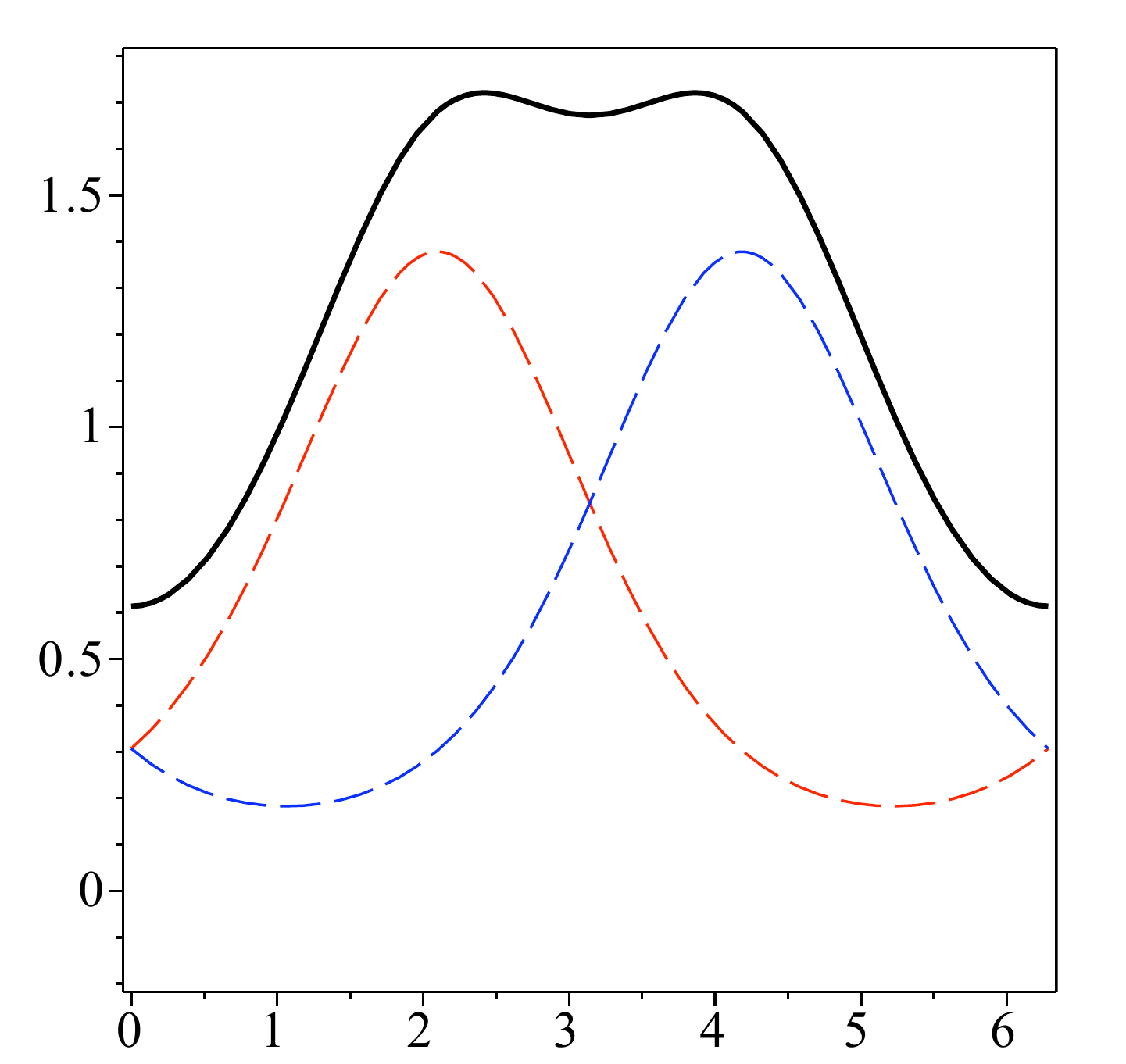}
  \caption{\label{fig_hat1}(Color online) The spectrum vs $\phi_{p}$ at \,$p_{t} = 1\,GeV$\, and $2 \, GeV$, for
  \,$dE/dx = 1\,GeV/fm$.\, The contribution to the spectrum mostly comes from a phonon at the ``cross'' in Fig.2.
  The red and blue dashed lines show contributions of the jets in the upper $(\alpha>0)$ and lower $(\alpha< 0)$ half plane in Fig.1.}
\end{figure}



\begin{figure}[t!]
  \vskip 0.3in
  \hskip -0.2in
  \includegraphics[width=8cm]{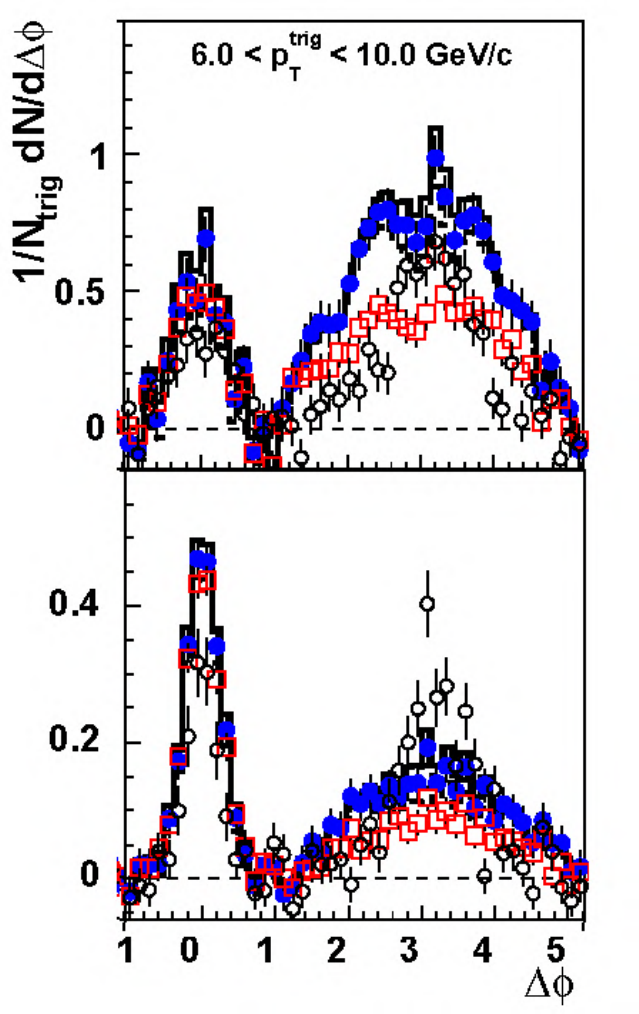}
  \caption{\label{fig_Exper}(Color online)  From \cite{STAR}.
   Background-subtracted azimuthal angle difference distributions for near-central 
   collisions (fraction of the total cross section 0-12\%). The associated particles 
   have the range of $p_T$ between $0.5 - 1$\,$GeV/c$ (upper figure), between 
   \,$1.5 - 2.5$\,$GeV/c$\, (lower figure), and the trigger particles have $p_T$ 
   ranging from \,$6.0$ to $10.0\,GeV/c$.\, The data for Au+Au collisions are shown 
   by the solid circles and for d+Au by the open circles. The rapidity range is $|\eta| <1$ 
  and as a result the rapidity difference is $|\Delta \eta | <2$. Open red squares show 
  results for a restricted acceptance of $|\Delta \eta| <0.7$. The solid and dashed 
  histograms show the upper and lower range of the systematic uncertainty due to the 
  $v_2$ modulation subtracted.}
\end{figure}

Finally, 
 we can carry out calculations of the pions' spectra including the sound wave perturbations, and do a comparison with the STAR's two-pion correlation functions. 

The effect of both components of the flow we called above $\delta u^{(1)},\delta u^{(2)}$
are calculated, for each $\alpha$, and then summed over. We remind that
the former effect is the effect of perturbed flow, normal to Mach cone, and the
latter is the effect of extra matter added by the wave, which flows
together with the rest of matter in the local radial flow direction. 
The results, for $p_t=1$ and $2 \, GeV/c$ are shown in Fig. \ref{fig_hat1}.
As shown there, the average over half plane produce a signal shifted in angle,
but in average together with another half-plane it creates a plateau-like
sum. As $p_t$ grows, it starts develop a double-hump structure
reminiscent of the original Mach cone predictions (but with a different angle). 

An example of hard-soft correlation function 
exhibited in Fig.\,\ref{fig_Exper} is from STAR publication \cite{STAR}.
The trigger $p_t=6..10\, GeV$, which we consider to be well in the hard hadron domain.
The open (black) circles correspond to dAu collision and are shown for comparison only, as in this case there is no matter for jets to interact with: they display narrow peak at $\phi=\pi$. The central  AuAu data are shown by (blue) filled circles and open (red)
squares, for two rapidity slices. Two pictures are for two bins
of the associate particle (see the captions). 

The overall shape and the width of the angular distribution seem to be reproduced
by the theory.

While comparing the two, one should however remember that:\\
(i) our calculation include only energy deposited into the media and the sound waves.
Sufficiently peripheral jets (with $\alpha\approx 90^o$) have punch-through
remnants of the jets which are not included. This component is increasing with
associate particle $p_t$, filling the place between the two humps in the lower picture.\\
(ii) theory calculation is done for pions having the same rapidity as the associate jet,
while the data are for certain  rapidity interval. However, as comparison
of the solid (blue) circles and open (red) squares shows,  the difference in shape
is not too strong.   This issue definitely need to be studied more, to separate
what we call ``cone" from the rapidity-independent ``circles".
\\
(iii) theory calculation have no fluctuations in $dE/dx$, and thus assume
completely deterministic scenario in which all jets stop at some fixed points, when its energy is all lost. Perhaps in reality fluctuations are not small, and there are punch-through jets at any $\alpha$.  

Completing the paper we emphasize that new data from LHC have
well reconstructed di-jet events, in which transverse momenta and rapidities
of both trigger and associate jets are detected. Soft particles angular distributions,
both in azimuth and rapidity, will obviously provide much better understanding
of the  issues involved. However current statistics of such events $\sim 1000$
is perhaps too small to address it. We will report calculations of the sound wave
contributions in such conditions elsewhere.

\vskip 1cm
\vskip .25cm {\bf Acknowledgments.}  This work was
supported in parts by the US-DOE grant DE-FG-88ER40388.


\section{Appendix A: The freezeout surface}

We approximate the  freezeout surface 
as the isoterm, with $T_f=120 \, MeV$. The
 parameterization of it is based on the results of other authors, namely
  Refs.\,\cite{hydro}, 
\cite{KolbHeinz:2002} and \cite{Knoll:2009}. 
The surface we use is parameterized as
\ba 
\tau(r)&=&-0.0011*r^5.995+0.0266*r^4.99762 \\ \nonumber 
&&-0.2367*r^4 +0.9772*r^3-1.9095*r^2\\ \nonumber  && +0.7*r+17.657
\ea

In \cite{hydro}, it 
was noted that the freezeout time changes only a little if one goes 
from SPS \,$(t_{f} = 10\,fm)$\, to RHIC \,$(t_{f} = 11\,fm)$.\, 
In \cite{KolbHeinz:2002}, the freezeout 
time for central collisions is calculated to be \,$\approx 12\,fm$\, 
\,(at $T_{f} = 130\,fm\,\,\mbox{for\,RHIC}$)
 In \cite{Knoll:2009}, the value obtained for
the freezeout time is \,$t_{f}=12\,fm/c$,\, however,
the temperature distributions can have sizable spreads in the
range of \,$\sim 100 - 160\,MeV$.\, In \cite{Lisa:2007}, having
the results of \cite{Eskola:2005}, it was pointed out that the
evolution time to kinetic freezeout - say until \,$T_{f}\approx
140\,MeV$\, is \,$t_{f}\sim 12-14\,fm/c$\, in both cases of
longitudinal and transverse dynamics; the timescale which is most
directly probed by femtoscopy. Nonetheless, from all these
scenarios, having a worked out combined scenario from \cite{hydro} 
and \cite{KolbHeinz:2002} at our disposal, we take the aforementioned  
values for the freezeout time and ``kinetic'' temperature as averaged 
values for our ansatz: namely \,$<t_{f}> \equiv t _{f} = 12\,fm$\, 
and  \,$<T_{f}> \equiv T_{f} =120\,fm$. Such a combined scenario
is based on the joint consideartion of these papers. 

Usually one separates ``kinetic" and ``chemical" freezeouts in 
which elastic and inelastic scattering rates are involved. Since different 
secondaries (pions, $K$ mesons, nucleons, $J/\psi$ particles, etc) in fact 
have quite different elastic cross sections, the ``kinetic" surfaces should 
be different for each species. We do not discuss all such complications 
in our work, and consider only one type of secondaries, the pions, the 
spectra of which are shown in the last section.

\section{Appendix B. Normalization of the Mach cone wave}

Suppose that the relative changes of the density $\rho$ and pressure $p$ in the medium of the fireball 
are much small as compared to their unperturbed values: \,$\delta p \ll p_{0}$\, and  \,$\delta\rho \ll \rho_{0}$ such that
\be %
p = p_{0} + \delta p\,\,\,\,\,\,\,\,\,\,\mbox{and}\,\,\,\,\,\,\,\,\,\,\rho = \rho_{0} + \delta\rho\,.
\label{eqn_app1}
\ee %
Suppose that this is also the case for the velocity of particles of the medium: $u\ll c_{s}$.

Then having the continuity equation and Euler's equation
\ba %
& & {{\partial \rho} \over {\partial t}} + \vec{\nabla} \rho\vec{u} = 0
\nonumber\\
& & {{\partial \vec{u}} \over {\partial t}}  + (\vec{u}\,\vec{\nabla})\vec{u} = -{\vec{\nabla}
 p \over \rho}\,,
 \label{eqn_app2}
\ea %
one can rewrite this system as
\ba %
& & {{\partial (\delta\rho)} \over {\partial t}} + \rho_{0}\,\vec{\nabla}\cdot \vec{u} = 0
\nonumber\\
& & {{\partial \vec{u}} \over {\partial t}}  + {\vec{\nabla}(\delta p) \over \rho_{0}} = 0\,.
\label{eqn_app3}
\ea %
If now we proceed to relativistic notations \,$\delta\rho=\delta\epsilon$\, and \,$\rho_{0} = \epsilon_{0} + p_{0}$,\, 
then the system (\ref{eqn_app3}) will have the following form:
\ba %
& & {{\partial (\delta\epsilon)} \over {\partial t}} + (\epsilon_{0}+p_{0})\vec{\nabla}\cdot \vec{u} = 0
\nonumber\\
& & (\epsilon_{0}+p_{0}){{\partial \vec{u}} \over {\partial t}}  + \vec{\nabla}(\delta p) = 0
\label{eqn_app4}\,.
\ea %
Actually, here the first equation is the energy conservation and the second one is the Newton's second law.

The small change in pressure $p$ is related to the small change in energy density $\epsilon$ by the adiabatic equation:
\be %
{{\delta p} \over {\delta\epsilon}} = \left( {{\partial p} \over {\partial\epsilon}} \right)_{\!s}\,\,\,\,\,\Rightarrow\,\,\,\,\,{\delta p} = c_{s}^{2}\,{\delta\epsilon}\,.
\label{eqn_app5}
\ee
The standard procedure with the system (\ref{eqn_app4}) leads to the wave equation both for $\delta p(x,y,z,t)$ and $\delta\epsilon(x,y,z,t)$ (see Ref.):
\be %
\nabla^{2}(\delta p) - {1 \over c_{s}^{2}}{{\partial^{2}(\delta p)} \over {\partial t^{2}}} = 0
\label{eqn_app6}
\ee
\be %
\nabla^{2}(\delta\epsilon) - {1 \over c_{s}^{2}}{{\partial^{2}(\delta\epsilon)} \over {\partial t^{2}}} = 0\,.
\label{eqn_app7}
\ee

We focus on the last equation which is a second order partial differential homogeneous equation. In case of a jet propagation which
is a source of sound waves, in the rhs of (\ref{eqn_app7}) there will be some function \,$f(x,y,z,t)$\, describing the source, and making 
the equation nonhomogeneous:
\be %
\nabla^{2}(\delta\epsilon) - {1 \over c_{s}^{2}}{{\partial^{2}(\delta\epsilon)} \over {\partial t^{2}}} = f(x,y,z,t)\,.
\label{eqn_app8}
\ee
Our goal is to solve the equation (\ref{eqn_app8}) which is carried out as follows. First we should find the solution of the homogeneous equation 
(\ref{eqn_app7}). In turn it can be rewritten in terms of cylindrical-coordinate wave equation such as
\ba %
& & {1 \over c_{s}^{2}}{{\partial^{2}(\delta\epsilon)} \over {\partial t^{2}}} =  
\nonumber\\
& &= {{\partial^{2}(\delta\epsilon)} \over {\partial r_{t}^{2}}} +
{1 \over r_{t}}{{\partial(\delta\epsilon)} \over {\partial r_{t}}} +
{1 \over r_{t}^{2}}{{\partial^{2}(\delta\epsilon)} \over {\partial \varphi^{2}}} +
{{\partial^{2}(\delta\epsilon)} \over {\partial x^{2}}}\,, 
\label{eqn_app9}
\ea %
where \,$\delta\epsilon \equiv \delta\epsilon(r_{t},\varphi,x,t)$,\, \,$r_{t}^{2} = y^{2} + z^{2}$,\, and the $\varphi$ is the azimuthal angle. 
The solution to the
cylindrical-coordinate wave equation can be found using the separation of variables:
\be %
\delta\epsilon(r_{t},\varphi,x,t) = R(r_{t})\,\Phi(\varphi)\,X(x)\,T(t)\,.
\label{eqn_app10}
\ee
Solving by this way one gets two final solutions in what follows.
\be %
\!\!\!\!\!\!\!\!\!\!
\delta\epsilon_{k_{t},k,n} = A_{k_{t},k,n}^{\pm}J_{n}(k_{t}r_{t})\,e^{\pm i\sqrt{k^{2} - k_{t}^{2}}\,x \pm ikc_{s}t \pm in\varphi}\,,
\label{eqn_app11}
\ee
\be %
\!\!\!\!\!\!\!\!\!\!
\delta\epsilon_{k_{t},k,n} = B_{k_{t},k,n}^{\pm}Y_{n}(k_{t}r_{t})\,e^{\pm i\sqrt{k^{2} - k_{t}^{2}}\,x \pm ikc_{s}t \pm in\varphi}\,,
\label{eqn_app12}
\ee
where $k_{t}$ and \,$k_{x}=\sqrt{k^{2} - k_{t}^{2}}$\, are the transverse and longitudinal momenta, and $J_{n}(k_{t}r_{t})$  and 
$Y_{n}(k_{t}r_{t})$ are the Bessel functions of the first and second kind. The $A_{k_{t},k,n}^{\pm}$ and $B_{k_{t},k,n}^{\pm}$
are some constants coming from solutions with the separate variables. The notation in the exponent of  (\ref{eqn_app11}) (or (\ref{eqn_app12})) 
means that there are eight combinations of the sum corresponding to positive and negative values of $x$, $k$ and $\varphi$. 
At azimuthal symmetry the term \,${1 \over r_{t}^{2}}{{\partial^{2}(\delta\epsilon)} \over {\partial \varphi^{2}}}$\, from 
(\ref{eqn_app9}) is dropped out which means that \,$n=0$\, in (\ref{eqn_app11}) and (\ref{eqn_app12}). Thus we can rewrite them as
\be %
\delta\epsilon_{k_{t},k} = A_{k_{t},k}^{\pm}J_{0}(k_{t}r_{t})\,e^{\pm i(\sqrt{k^{2} - k_{t}^{2}}\,x + kc_{s}t)}\,,
\label{eqn_app13}
\ee
\be %
\delta\epsilon_{k_{t},k} = B_{k_{t},k}^{\pm}Y_{0}(k_{t}r_{t})\,e^{\pm i(\sqrt{k^{2} - k_{t}^{2}}\,x + kc_{s}t)}\,,
\label{eqn_app14}
\ee     
and (\ref{eqn_app8}) as
\be %
 {{\partial^{2}(\delta\epsilon)} \over {\partial t^{2}}} - c_{s}^{2}\nabla^{2}(\delta\epsilon)  = f(t,x,y,z)\,.
\label{eqn_app15}
\ee
In this case in the exponent of  (\ref{eqn_app13}) (or (\ref{eqn_app14}))  there are four combinations of the sum corresponding to
 positive and negative values of $x$ and $k$. 

We solve this last equation using (\ref{eqn_app13}) and/or (\ref{eqn_app14}). Suppose we use (\ref{eqn_app13}). Then the operators in 
the l.h.s. of  (\ref{eqn_app15}) take the following forms:
\be %
\!\!\!\!\!\!\!\!\!\!
 {{\partial^{2}(\delta\epsilon)} \over {\partial t^{2}}}  =  -\omega^{2} \left[ A_{k_{t},k}^{\pm}J_{0}(k_{t}r_{t})\,e^{\pm i(\sqrt{k^{2} - k_{t}^{2}}\,x + \omega t)} \right]\,,
\label{eqn_app16}
\ee
\ba %
& & - c_{s}^{2}\nabla^{2}(\delta\epsilon)  = 
\nonumber\\
& & = c_{s}^{2}(k_{t}^{2} + k_{x}^{2}) \,\times
\nonumber\\
& & \times \left[ A_{k_{t},k}^{\pm}J_{0}(k_{t}r_{t})\,e^{\pm i(\sqrt{k^{2} - k_{t}^{2}}\,x + \omega t)} \right]\,.
\label{eqn_app17}
\ea
By means of these two expressions one can rewrite the solution of (\ref{eqn_app15}) in the $\omega$ and $k$ space, such as
\be %
\left[ -\omega^{2} +c_{s}^{2}(k_{t}^{2} + k_{x}^{2}) \right]\phi_{\omega k} = f_{\omega k} \equiv f(\omega,k)\,.
\label{eqn_app18}
\ee
The $f(\omega,k)$ is determined by the Fourier transform of \,$f(x,y,z,t)$\, which one can choose as a function of the
form \,$(1/ct_{f})\delta(\pm x - ct)\delta(y)\delta(z)$:
\ba %
& & f(\omega,k) = 
\nonumber\\
& & = \int dt\,dx\,dy\,dz\,e^{\pm i(\sqrt{k^{2} - k_{t}^{2}}\,x + \omega t)} \,\times
\nonumber\\
& & \times\,\,{1 \over ct_{f}}\,\delta(\pm x - ct)\delta(y)\delta(z)\,,
\label{eqn_app19}
\ea    
The $c$ is the speed of light. From  (\ref{eqn_app18}) the $\phi_{\omega k}$ is represented as
\be %
\phi_{\omega k} ={f(\omega,k)  \over c_{s}^{2}k^{2} - \omega^{2}}\,.
\label{eqn_app20}
\ee     
Plugging the result of the integral of (\ref{eqn_app19}) into this equation we obtain
\ba %
& & (\phi_{\omega k})_{+} = 
\nonumber\\
& & = \left( {2\pi \over c^{2}t_{f}} \right) \left({1 \over c_{s}^{2}k^{2} - \omega^{2}} \right)
\delta\!\left( {\omega \over c} - k_{x} \right)\,,
\nonumber\\
\nonumber\\
\nonumber\\
& &  (\phi_{\omega k})_{-} = 
\nonumber\\
& & = \left( {2\pi \over  c^{2}t_{f}} \right) \left({1 \over c_{s}^{2}k^{2} - \omega^{2}} \right)
\delta\!\left( {\omega \over c} - (-k_{x}) \right)\,.
\label{eqn_app21}
\ea %
The \,$(\phi_{\omega k})_{+}$\, is for positive $k_{x}$, and the \,$(\phi_{\omega k})_{-}$\, is for negative $k_{x}$.
Carring out the inverse Fourier transform of $\phi_{\omega k}$ along with (\ref{eqn_app21}) we get
\ba %
& &  [\phi(t,x,y,z)]_{+} = [\delta\epsilon(t,x,y,z)]_{+} =
\nonumber\\
& & = \int {d\omega\,d^{3}k \over (2\pi)^{4}}\,e^{\mp i(\omega t - k_{x}x)} (\phi_{\omega k})_{+}\,\,,
\nonumber\\
\nonumber\\
\nonumber\\
& &  [\phi(t,x,y,z)]_{-} = [\delta\epsilon(t,x,y,z)]_{-} = 
\nonumber\\
& & = \int {d\omega\,d^{3}k \over (2\pi)^{4}}\,e^{\mp i(\omega t -  (-k_{x})x)} (\phi_{\omega k})_{-}\,,
\label{eqn_app22}
\ea %
which consecutively results in 
\ba %
& &  [\delta\epsilon(t,x,y,z)]_{+} = 
\nonumber\\
& & = {1 \over (2\pi)^{3} ct_{f}}\int e^{\mp i(k_{x}ct - (k_{x}x + k_{y}y + k_{z}z))}\,\times 
\nonumber\\
& & \times\,{dk_{x}\,dk_{y}\,dk_{x} \over c_{s}^{2} (k_{x}^{2} + k_{y}^{2} + k_{z}^{2}) - c^{2}k_{x}^{2}}\,,
\nonumber\\
\nonumber\\
\nonumber\\
& & [\delta\epsilon(t,x,y,z)]_{-} = 
\nonumber\\
& & = {1 \over (2\pi)^{3} ct_{f}}\int e^{\mp i(-k_{x}ct + (k_{x}x + k_{y}y + k_{z}z))}\,\times
\nonumber\\
& & \times\,{dk_{x}\,dk_{y}\,dk_{x} \over c_{s}^{2} (k_{x}^{2} + k_{y}^{2} + k_{z}^{2}) - c^{2}k_{x}^{2}}\,,
\label{eqn_app23}
\ea %
then
\ba %
\Rightarrow & & [\delta\epsilon(t,x,y,z)]_{+} = 
\nonumber\\
& &= {1 \over (2\pi)^{3} ct_{f}}\int dk_{x}\,e^{\mp i(k_{x}ct - k_{x}x)}\,\times
\nonumber\\
& & \times\,\int dk_{t}\,d\varphi\,{k_{t}\,e^{\mp ik_{t}r_{t}\cos{\varphi}} \over c_{s}^{2} (k_{x}^{2}  + k_{t}^{2}) - c^{2}k_{x}^{2}}\,,
\nonumber\\
\nonumber\\
\nonumber\\
& & [\delta\epsilon(t,x,y,z)]_{-} = 
\nonumber\\
& & = {1 \over (2\pi)^{3} ct_{f}}\int dk_{x}\,e^{\mp i(- k_{x}ct + k_{x}x)}\,\times
\nonumber\\
& & \times\,\int dk_{t}\,d\varphi\,{k_{t}\,e^{\pm ik_{t}r_{t}\cos{\varphi}} \over c_{s}^{2} (k_{x}^{2}  + k_{t}^{2}) - c^{2}k_{x}^{2}}\,,
\label{eqn_app24}
\ea %

\ba %
\Rightarrow & & [\delta\epsilon(t,x,y,z)]_{+} =
\nonumber\\
& & =  {1 \over (2\pi)^{2} ct_{f}}\int dk_{x}\,e^{\mp i(k_{x}ct - k_{x}x)}\,\times
\nonumber\\
& & \times\,\int dk_{t}\,{(k_{t}/ c_{s}^{2})\,J_{0}(k_{t}r_{t}) \over k_{t}^{2} + [(c_{s}^{2}-c^{2})/c_{s}^{2}]k_{x}^{2}}\,,
\nonumber\\
\nonumber\\
\nonumber\\
& & [\delta\epsilon(t,x,y,z)]_{-} = 
\nonumber\\
& & = {1 \over (2\pi)^{2} ct_{f}}\int dk_{x}\,e^{\mp i(- k_{x}ct + k_{x}x)}\,\times
\nonumber\\
& & \times\,\int dk_{t}\,{(k_{t}/ c_{s}^{2})\,J_{0}(k_{t}r_{t}) \over k_{t}^{2} + [(c_{s}^{2}-c^{2})/c_{s}^{2}]k_{x}^{2}}\,.
\label{eqn_app25}
\ea %
In the integral over $k_{t}$ the first order pole is
\be %
 k_{t} = \pm i\sqrt{{c_{s}^{2}-c^{2} \over c_{s}^{2}}}\,k_{x}\,.
\label{eqn_app26}
\ee
Hereby, we further have
\ba %
& & [\delta\epsilon(t,x,y,z)]_{+} = 
\nonumber\\
& & = {(2\pi i) \over (2\pi)^{2} ct_{f}}\int dk_{x}\,e^{\mp i(k_{x}ct - k_{x}x)}\,\times
\nonumber\\
& & \times\,\left[ {(k_{t}/ c_{s}^{2}) J_{0}(k_{t}r_{t}) \over 2k_{t}} \right]_{ k_{t} = \pm i\sqrt{{c_{s}^{2}-c^{2} \over c_{s}^{2}}}\,k_{x}}\,,
\nonumber\\
\nonumber\\
\nonumber\\
& & [\delta\epsilon(t,x,y,z)]_{-} = 
\nonumber\\
& & = {(2\pi i) \over (2\pi)^{2} ct_{f}}\int dk_{x}\,e^{\mp i(- k_{x}ct + k_{x}x)}\,\times
\nonumber\\
& & \times\,\left[ {(k_{t}/ c_{s}^{2}) J_{0}(k_{t}r_{t}) \over 2k_{t}} \right]_{ k_{t} = \pm i\sqrt{{c_{s}^{2}-c^{2} \over c_{s}^{2}}}\,k_{x}}\,,
\label{eqn_app27}
\ea %

\ba %
\Rightarrow & & [\delta\epsilon(t,x,y,z)]_{+} = 
\nonumber\\
& & = {i \over 4\pi c_{s}^{2} ct_{f}}\int dk_{x}\,e^{\mp i(k_{x}ct - k_{x}x)}\,\times
\nonumber\\
& & \times\,J_{0}\!\!\left( \pm i\sqrt{c_{s}^{2} - c^{2} \over c_{s}^{2}}\,r_{t}k_{x} \right)\,,
\nonumber\\
\nonumber\\
\nonumber\\
& &[\delta\epsilon(t,x,y,z)]_{-} = 
\nonumber\\
& & = {i \over 4\pi c_{s}^{2} ct_{f}}\int dk_{x}\,e^{\mp i(-k_{x}ct + k_{x}x)}\,\times
\nonumber\\
& & \times\,J_{0}\!\!\left( \pm i\sqrt{c_{s}^{2} - c^{2} \over c_{s}^{2}}\,r_{t}k_{x} \right)\,.
\label{eqn_app28}
\ea %

Ultimately, the integration of two integrals in (\ref{eqn_app28}) gives the following solutions for the function $\delta\epsilon(t,x,y,z)$:
\ba %
& & [\delta\epsilon(t,x,y,z)]_{+} =
\nonumber\\
& & = 0\,,
\nonumber\\
& & = - {1 \over 2\pi c_{s}ct_{f}}{1 \over\sqrt{ c_{s}^{2}(x - ct)^{2} + (c^{2} - c_{s}^{2})r_{t}^{2}}}\,,
\nonumber\\
& & = {1 \over 2\pi c_{s}ct_{f}}{1 \over \sqrt{c_{s}^{2}(x - ct)^{2} + (c^{2} - c_{s}^{2})r_{t}^{2}}}\,,
\nonumber\\
& & = 0\,,
\label{eqn_app29}
\ea %
and
\ba %
 & & [\delta\epsilon(t,x,y,z)]_{-}  = 
\nonumber\\
 & & = 0\,,
\nonumber\\
 & & = {1 \over 2\pi c_{s}ct_{f}}{1 \over\sqrt{ c_{s}^{2}(x - ct)^{2} + (c^{2} - c_{s}^{2})r_{t}^{2}}}\,,  
\nonumber\\
 & & = - {1 \over 2\pi c_{s}ct_{f}}{1 \over \sqrt{c_{s}^{2}(x - ct)^{2} + (c^{2} - c_{s}^{2})r_{t}^{2}}}\,,  
\nonumber\\
 & & = 0\,.  
\label{eqn_app30}
\ea %
Combining the solutions which are in (\ref{eqn_app29}) and (\ref{eqn_app30}) we obtain
\ba %
& & [\delta\epsilon(t,x,y,z)]_{+} = [\delta\epsilon(t,x,y,z)]_{-} = 
\nonumber\\
& & = \pm{1 \over 2\pi c_{s}ct_{f}}{1 \over\sqrt{ c_{s}^{2}(x - ct)^{2} + (c^{2} - c_{s}^{2})(y^{2} + z^{2})}}\,.
\nonumber\\
& & = 0\,.
\label{eqn_app31}
\ea %
Generally, the function \,$f(t,x,y,z)$\, in equation  (\ref{eqn_app15}) can include some constant $A$ which has a dimension
of $GeV/fm$. Taking also into account this constant we have
\ba %
& & \delta\epsilon(t,x,y,z) = 
\nonumber\\
& & = {1 \over 2\pi c_{s}ct_{f}}{A \over\sqrt{ c_{s}^{2}(x - ct)^{2} + (c^{2} - c_{s}^{2})(y^{2} + z^{2})}}\,.
\label{eqn_app32}
\ea %
One can make this constant $A$ equal to \,$dE/dx$\, which is the loss of the jet energy per unit path length.


\end{document}